\documentclass[%
 reprint,
 amsmath,amssymb,
 aps,
]{revtex4-2}

\usepackage{graphicx}
\usepackage{dcolumn}
\usepackage{bm}

\usepackage{enumitem} 
\usepackage{subfigure}
\usepackage{float}
\usepackage{color}
\usepackage{ulem}
\usepackage{amsmath}
\usepackage{physics}

\def\cD{{\mathcal D}}
\def\cE{{\mathcal E}}
\def\cS{{\mathcal S}}
\def\hH{{\hat H}}
\def\vx{\mathbf{x}}

\newcommand\COMMENTED[1] {}

\begin{document}


\title{Implementation of correlated trial wave functions  \\
in auxiliary-field quantum Monte Carlo by stochastic sampling}

\title{Implementing advanced trial wave functions  \\
in fermion quantum Monte Carlo via stochastic sampling}

\author{Zhi-Yu Xiao}
\affiliation{%
Institute of Physics, Chinese Academy of Sciences, P.O. Box 603, Beijing 100190, China
}%
\email{zxiaoMain@outlook.com}

\author{Zixiang Lu}
\affiliation{%
University of Illinois Urbana-Champaign, Champaign, IL, USA
}%

\author{Yixiao Chen}
\affiliation{%
Princeton University, Princeton, NJ, USA
}%

\author{Tao Xiang}
\affiliation{%
Institute of Physics, Chinese Academy of Sciences, P.O. Box 603, Beijing 100190, China
}%

\author{Shiwei Zhang}
\affiliation{%
Center  for  Computational  Quantum  Physics,  Flatiron  Institute,  New  York,  NY  10010,  USA
}%
\email{szhang@flatironinstitute.org}

\date{\today}

\begin{abstract}

We introduce an efficient approach to implement correlated many-body trial wave functions in auxiliary-field quantum Monte Carlo (AFQMC).  To control the sign/phase problem in AFQMC, a constraint is derived from an exact gauge condition but is typically imposed approximately through a trial wave function or trial density matrix, whose quality can affect the accuracy of the method. Furthermore, the trial wave function can also affect the efficiency through importance sampling. The most natural form of the trial wave function has been single Slater determinants or their linear combinations. More sophisticated forms, for example, with the inclusion of a Jastrow factor or other explicit correlations, 
have been challenging to use and their implementation is  often assumed to require a quantum computer.
In this work, we demonstrate that a large class of correlated wave functions, written in the general form of multi-dimensional integrals over hidden or auxiliary variables times Slater determinants, can be implemented 
as trial wave function
by coupling the random walkers to 
a generalized Metropolis sampling.
We discuss the fidelity of AFQMC with stochastically sampled trial wave functions, which are relevant to both quantum and classical algorithms. We 
illustrate the method and show that an efficient implementation can be achieved which
preserves the low-polynomial computational scaling of AFQMC. We
test our method in molecules under bond stretching and in transition metal diatomics. Significant improvements are seen in both accuracy and efficiency over typical trial wave functions, and the method yields total ground-state energies systematically within chemical accuracy. 
The method can be useful for incorporating other advanced wave functions, for example, neural quantum state wave functions optimized from machine learning 
techniques, or for other forms of fermion quantum Monte Carlo.

\end{abstract}

\pacs{Valid PACS appear here}

\maketitle

\section{\label{sec:introduction}Introduction}
The study of interacting quantum many-body systems is a major challenge across diverse fields such as condensed matter physics, nuclear physics, cold atoms physics, quantum chemistry, and materials science. Due to the high degree of complexity and the interplay between the various degrees of freedom in quantum many-body systems, 
no universal approach exists at present that can fully address the complexity of interacting quantum systems while providing systematically accurate results across a wide range of many-body models and materials. Different methods have been developed, each specializing in particular types of systems or aspects of the same system. Continued development of more general and precise numerical techniques is crucial to overcoming the challenges, and to accelerating progress in understanding and predicting the properties of interacting quantum systems. 

Quantum Monte Carlo (QMC) methods \cite{becca_QMC_correlated_2017, 
foulkes_solids_2001, RevModPhys_nuclear_2015,lecturenotes-2019}, as one of the main many-body techniques, are widely used in the study of quantum many-body systems. In equilibrium state QMC, the many-body wave function or density matrix is sampled within a chosen basis. Depending on factors such as the form of the Hamiltonian, the computation’s objective (e.g., finite-temperature vs. ground-state properties), and the chosen basis, QMC methods can vary significantly. These variations give rise to different "flavors" of Monte Carlo sampling, each tailored to specific needs. For most non-trivial quantum systems, achieving adequate sampling efficiency requires extensive customization of the Monte Carlo technique.

Implementations of QMC  
are typically based on one of two primary sampling algorithmss: Markov Chain Monte Carlo (MCMC) or branching random walks (BRW). 
A QMC calculation then will follow one of these frameworks. In some cases, the distinction between the two methods is primarily a matter of efficiency or convenience, while in others, the difference is more fundamental. For instance, in auxiliary-field quantum Monte Carlo (AFQMC), MCMC is often employed within a path-integral formalism when there is no sign problem, or when an acceptable average sign or phase can be maintained to yield meaningful results. This approach is commonly used in lattice quantum chromodynamics \cite{AFQMC_L_QCD} and condensed matter systems \cite{QMC_DQMC}, for example in 
determinantal quantum Monte Carlo (DQMC). When the sign or phase problem becomes more severe, a class of algorithms referred to as constrained path \cite{QMC_Zhang_Constrained_1997,zhang1999finite} or phaseless \cite{AFQMC_Zhang-Krakauer-2003-PRL} AFQMC are
designed to control these issues by imposing sign or gauge conditions.
These methods require the use of BRW 
along the imaginary-time paths 
because the 
introduction of constraints 
renders the MCMC sampling exponentially inefficient to achieve ergodicity \cite{lecturenotes-2019}.

AFQMC with constraints have enabled a wide range of applications where sign-problem-free computations are not feasible, including strongly correlated models in condensed matter \cite{Bo_Xiao_Stripe_Hubbard, Hubbard_SC_HAOXU}, bulk solids \cite{Mario_hydrogen_chain,BM_model}, and quantum chemistry \cite{WIREs_Mario,Joonho_chemistry}. 
These methods are approximate, and their accuracy can be affected by the quality of trial wave function or trial density matrix. 
Typically AFQMC employs Slater determinant wave functions or some variants of 
the Slater determinant, for example, orthogonal or non-orthogonal linear combinations 
of Slater determinants \cite{Hao_Some_recent_developments,WIREs_Mario, Morales_MD_AFQMC, SHCI_AFQMC},
symmetry-projected mean-field solutions \cite{Hao_symmetry},
projected BCS \cite{PRA_2011_2DFGs_Carlson, Vitali_Calculating_2019} or pseudo-BCS \cite{Zxiao_2020} as trial wave functions.  In a variety of correlated fermion systems, this framework has been 
shown to be systematic across multiple benchmark studies
\cite{Hubbard_benchmark_2015,Joonho_chemistry, Mario_hydrogen_chain}.
For certain systems, however, 
it has been necessary to go beyond simple Slater determinants, 
by use of multi-determinant trial wave functions 
\cite{simons_material_2020, transition_metals_Shee,Hao_Some_recent_developments,SHCI_AFQMC, Ankit_CISD}.
In extended systems, scaling up the trial wave function in a configuration interaction type of approach does not lead to size-consistency. In strongly correlated models, the most effective approach for improving the constraint has been 
through the use of self-consistency \cite{Mingpu_SC,Yuan-Yao_FT_AFQMC,Hubbard_SC_HAOXU}.

Using other more correlated trial wave functions has been more challenging. The
difficulty can be summarized in simple terms as that of evaluating the overlap of a Slater determinant (a random walker in the BRW), $|\phi\rangle$, with a general trial wave function $|\Psi_T\rangle$ which is in an explicitly correlated
 form, for example, a Slater determinant times a Jastrow function (Slater-Jastrow), coupled-clusterwave function, or matrix product state, etc.
There have been various attempts to overcome this difficulty \cite{Chang_MD_AFQMC, Ankit_CISD, MPS_AFQMC,yu2025_MPS_AFQMC}. 
An interesting proposal, under the name QC-AFQMC \cite{Huggins2022_AFQMC_quantum_computer}, of using a quantum computer in the
evaluation of the overlap $\langle\Psi_T|\phi\rangle$ has garnered considerable 
attention~\cite{QC_AFQMC_Huang_2024, Reduced_Quantum_Resources_AFQMC}.
However, a general algorithm is still not available which allows the use of sufficiently general, 
size-extensive, correlated trial wave functions in 
large systems with demonstrated 
computational scalability and high fidelity.

In this work, we 
show that it is possible to realize an efficient implementation of a large class of correlated trial wave functions in AFQMC with ``classical'' algorithms. A key pivot is to consider ratios of overlaps rather than the overlaps themselves.
In AFQMC  overlaps only appear in ratios,
$\langle\Psi_T|\phi'\rangle/\langle\Psi_T|\phi\rangle$, where  $|\phi'\rangle$ and  $|\phi\rangle$ are connected by one step in the random walk. The ratio drives importance sampling, and its sign or phase imposes the gauge condition that controls the sign or phase problem. The ratio is well-behaved, changes slowly, and scales gracefully with system size in the simulation.
In contrast,
even in a calculation of modest system size, 
the overlap itself can vary by 50 orders of magnitude \cite{Hao_infinite_variance}.
Thus any calculation based on direct evaluations of the overlap itself 
would have fundamental difficulty 
to scale up properly with system size. Even if the overlap could be measured, 
an important question would be how the fidelity of the constraint depends on
any noise incurred in the measurement, 
for example, from stochastic sampling. 

We realize the evaluation of the ratios by stochastically sampling the trial wave function.
More specifically, we show that any $|\Psi_T\rangle$ written in the form 
$|\Psi_T\rangle = \int p_T(Y)\,\hat {\mathcal B}_T(Y)\,dY\,|\phi_T\rangle$ (where 
$|\phi_T\rangle$ is a single (or finite number of) Slater determinant, $Y$ is a multi-dimensional variable, $p_T(Y)$ is a probability density function, and $\hat {\mathcal B}_T(Y)$ consists of exponentials of one-body operators) can be used as a trial wave function in AFQMC. Examples of this form include
Slater-Jastrow, coupled-cluster with singles and doubles (CCSD) \cite{CC_chem_origin}, or certain neural quantum 
state forms with hidden variables, such as that from variational AFQMC \cite{VAFQMC_Sorella,HAFQMC,VAFQMC_Ryan}. 

The algorithm takes the form of BRWs coupled to MCMC, with each random walker in AFQMC tethered to a number of MCMC paths in $Y$-space which are dynamically updated as the 
random walk proceeds. 
We show that the ratio can be estimated by a small number of samples for 
$\langle\Psi_T|$ to achieve sufficient fidelity in the constraint and high accuracy in the 
importance sampling. Crucial to the algorithm is that the samples are not static or frozen but rather move by MCMC as the random walkers evolve. 
This is also to be contrasted with an ``on-demand'' approach which provides samples from 
the trial wave function without any ``prior'' or ``conditioning'' from the current 
position of the walker, $|\phi\rangle$, which would suffer from large fluctuations 
from random noise and would not scale properly with system size.

The method paves the way for many types of correlated trial wave functions
to be employed in AFQMC. 
Our specific choice in this work has been a class of wave functions 
that can be represented as   
multi-dimensional integrals times Slater determinants, which can be obtained 
 variationally. As we show, even the simplest wave functions in this class can enable
significant improvements in accuracy in strongly correlated molecules. 
Our results are also relevant to much of the discussion 
surrounding the discussion on quantum-classical hybrid implementations \cite{Huggins2022_AFQMC_quantum_computer,Reduced_Quantum_Resources_AFQMC} 
of AFQMC. 
This paper introduces the conceptual framework for our method,  describes the algorithm for integrating MCMC into the BRW framework of AFQMC, presents details for practical implementation, demonstrates the effect in prototypical systems (N$_2$ bond stretching, transition metal oxide molecules), and discusses scaling and efficiency considerations. 
Our approach can be generalized to many other types of trial wave functions, for example neural quantum states (NQS) in orbital space, or to other
flavors of QMC. 

The rest of this paper is organized 
as follows. In Sec.~\ref{sec:Prilimilaries}, we provide as background 
an overview of the AFQMC formalism, brief discussion on the correlated trial wave functions we use, and an introduction to MCMC and BRW samplings in the context of auxiliary-fields 
to facilitate their integration. 
In Sec.~\ref{sec:AFQMC_Metro}, we introduce the formalism for implementing the 
new method, and provide an outline of the algorithm to help the reader with
implementations. In Sec.~\ref{sec:Results}, we present illustrative results 
in atoms/molecules, before concluding and making a few general remarks 
in Sec.~\ref{sec:summary}.
Additional details are provided in the APPENDIX.

\section{\label{sec:Prilimilaries} Background}

\subsection{\label{sec:Prilimilaries_overview} Preliminaries}
In this section, we introduce preliminaries 
and notations which will facilitate the discussions below 
of AFQMC, the trial wave functions we consider, and the new algorithm.
Any many-fermion 
Hamiltonian with one- and two-body terms 
can be written in second quantization 
as (suppressing spin index for brevity)
\begin{eqnarray}
{\hat H} = {\hat H_1}+{\hat H_2} 
= \sum_{i,j}^N {T_{ij} c_i^\dagger c_j}
   + {1 \over 2}
\sum_{i,j,k,l}^N {V_{ijkl} c_i^\dagger c_j^\dagger c_k c_l}\,,
\label{eq:H}
\end{eqnarray}
where $N$ is the size of the one-particle basis set.
For quantum chemistry basis sets (Gaussian type orbitals, or GTOs), the matrix elements 
are conveniently available and are all real. 

The exponential of any one-body operator  has a special role throughout this discussion, thanks to Thouless theorem:
\begin{equation}
{\rm exp}\big(\sum_{ij}c_i^\dagger U_{ij}c_j\big)\,|\phi\rangle=|\phi'\rangle\,,
\label{eq:spo}
\end{equation}
where $|\phi\rangle$ and $|\phi'\rangle$ are Slater determinants. The action of any operator which is the exponential of a one-body operator on a 
Slater determinant leads to another Slater determinant. 
The orbitals of $|\phi'\rangle$ can be 
easily obtained by matrix products involving the matrix elements of the one-body operator, $\{U_{ij}\}$, and the orbitals of $|\phi\rangle$ 
\cite{lecturenotes-2019}. We will often denote the exponential 
of one-body operators by $\hat B$ or $\hat {\mathcal B}$. 
For example, $e^{-\tau \hat H_1}$ is in this form (where $\tau$ is a number).

The exponential of a two-body operator can be written as 
a many-dimensional integral over the exponential one-body operators~\cite{lecturenotes-2019}. 
The two-body operator can be expressed in 
``Monte Carlo form" \cite{Hao_Some_recent_developments}:
\begin{equation}
\hat{H_2}\doteq \frac{1}{2}\sum^\Gamma_\gamma \hat{L}_\gamma ^2\,,
\label{eq:H_MC}
\end{equation}
where $\hat{L}_\gamma$ denote sets of one-body operators, whose matrix elements 
depend on details of the ``factorization" transformation
(a modified Cholesky \cite{Cholesky_dec}, density-fitting \cite{GPU_AFQMC_James}, etc) but 
are explicitly specified from $\{V_{ijkl}\}$, 
and $\Gamma\sim {\mathcal O}(10)\,N$.  
Applying a Hubbard-Stratonovich transformation \cite{HS_transformation_H, HS_transformation_S} we can obtain
\begin{equation}
e^{-\tau \hat H_2 }\doteq  \int\mathrm{d}\textbf{x}\, p(\textbf{x})\,\hat{B}_V(\textbf{x})\,,
\label{eq:HS-V}
\end{equation}
where $\textbf{x}$, called auxiliary fields, denotes a series of $\{x_\gamma\}$ with $x_\gamma \in \mathbb{R}$, and
$p(\textbf{x})= \prod^\Gamma_\gamma \frac{1}{\sqrt{2\pi}}e^{-x_\gamma^2/2}$
and related one-body propagator 
$\hat{B}_V(\textbf{x})=\exp(\sum_\gamma x_\gamma\sqrt{-\tau}\hat{L}_\gamma )$. (In $\hat B_V$ the subscript $V$  indicates that it came from $\{V_{ijkl}$.) 

Putting the above together, the short imaginary-time propagator can be written as 
\begin{equation}
e^{-\tau \hat{H} } \doteq \int\mathrm{d}\textbf{x}\, p(\textbf{x})\,\hat{B}(\textbf{x})\,,
\label{eq:HS}
\end{equation}
where
$\hat{B}(\textbf{x})=e^{-\tau\hat{T}/2}\,\hat B_V(\textbf{x})\,e^{-\tau\hat{T}/2}$. This provides the basic ingredient for AFQMC.

The above discussion is more general. For example, 
$e^{\hat T_1}$, where $\hat T_1$ denotes single excitations,
satisfies Eq.~(\ref{eq:spo}) as well; similarly for any mean-field operators. Two-body operators, for example double excitations $\hat T_2$ in coupled cluster, or the exponents in 
two-body Jastrow factors $\hat J_2$, can be treated as in 
Eq.~(\ref{eq:H_MC}) so that $e^{\hat T_2}$ or $e^{\hat J_2}$ 
is written in the form of Eq.~(\ref{eq:HS-V}).

\subsection{\label{sec:Prilimilaries_AFQMC} Outline of standard AFQMC}

We outline the AFQMC method in this section, building on the 
notation above and setting up the discussions on algorithms below.
In this paper, we focus on ground-state calculations, although our discussion can be easily generalized 
to finite temperature AFQMC \cite{zhang1999finite,Yuan-Yao_FT_AFQMC}. To reach the ground state, AFQMC 
simulates an ``imaginary-time evolution": 
\begin{equation}
|\Psi_{0}\rangle\propto \underset{\beta \to \infty}{\lim} e^{-\beta\hat{H}}|\Psi_I\rangle\,,
\label{eq:imaginary_time_evolve}
\end{equation}
which propagates the given initial state $|\Psi_I\rangle$ to the ground state $|\Psi_0\rangle$ of the Hamiltonian $\hat{H}$, if their overlap $\langle\Psi_{0}|\Psi_I\rangle$ is non-zero. 
Numerically, the propagation is performed with ``short-time" steps: 
\begin{equation}
e^{- \beta\hat{H} }  =   (e^{-\tau \hat{H} })^n\,,
\end{equation}
where $n= \beta/\tau $ denotes the number of time slices, with
a sufficiently small 
time-step $\tau$. . 

Throughout this paper, we will use $|\phi\rangle$ (with sub- and superscripts as needed) to denote single Slater determinant wave functions, and $|\Psi\rangle$ to denote correlated many-body wave functions beyond a single Slater determinant. 

In AFQMC, to control the sign/phase problem, an 
importance-sampled BRW algorithm is used to simulate the ``imaginary-time evolution" in Eq.~\ref{eq:imaginary_time_evolve}~\cite{QMC_Zhang_Constrained_1997,AFQMC_Zhang-Krakauer-2003-PRL}, by turning it into an
iterative propagation of 
\begin{equation}
e^{-\tau \hat{H} }|\Psi^i\rangle \rightarrow |\Psi^{i+1}\rangle\,,
\end{equation}
with a single Slater determinant, a linear combination of Slater determinants, or determinants sampled from the trial wave function, to initialize $|\Psi^{0}\rangle=|\Psi_I\rangle$. 
The wave function 
$|\Psi^i\rangle$ at step $i$ of the BRW is represented by an ensemble of random walkers, $\{|\phi_k^i\rangle, W_k^i\}$:
\begin{equation}
 |\Psi^i\rangle
\propto \sum_k W^i_k\,\frac{|\phi^i_k \rangle}{\langle \Psi_T|\phi^i_k \rangle}\,, 
\label{eq:WF-imp}
\end{equation}
where $k$ labels the walker, and $W^i_k$ is the weight of walker $k$, and the denominator comes from importance sampling 
\cite{lecturenotes-2019}.
This is achieved by taking a dynamic shift $\overline{\textbf{x}}$ in the integration over auxiliary fields in Eq.~(\ref{eq:HS}):
\begin{equation}
e^{-\tau \hat{H} }=  \int\mathrm{d}\textbf{x}\, p(\textbf{x}- \overline{\textbf{x}})\,\hat{B}(\textbf{x}- \overline{\textbf{x}})\,,
\label{eq:HS-shifted}
\end{equation}
with which each step of propagation can be performed as:
\begin{equation}
e^{-\tau \hat{H} }\,
\sum_k W^{i}_k \frac{|\phi^{i}_k \rangle}{\langle \Psi_T|\phi^{i}_k \rangle} 
 \rightarrow 
 \sum_k W^{i+1}_k \frac{|\phi^{i+1}_k \rangle}{\langle \Psi_T|\phi^{i+1}_k \rangle}\,
\label{eq:prop-imp}
\end{equation}
that advances the walker 
\begin{equation}
\begin{aligned}
\hat{B}(\textbf{x} - \overline{\textbf{x}}_k^i)| \phi_k^i \rangle
\rightarrow | \phi_k^{i+1} \rangle\,,
\label{eq:walker_prop}
\end{aligned}
\end{equation}
and assigns the weight
\begin{equation}
\begin{aligned}
I(\textbf{x}, \overline{\textbf{x}}_k^i, \phi_k^i)\,W_k^i \rightarrow W_k^{i+1}
\label{eq:weight_prop}
\end{aligned}
\end{equation}
with
\begin{equation}
\begin{aligned}
I(\textbf{x}, \overline{\textbf{x}}^{i}_k, \phi^{i}_k) = \frac{p(\textbf{x}-\overline{\textbf{x}}^{i}_k)}{p(\textbf{x})}\frac{\langle \Psi_T| \hat{B}(\textbf{x} - \overline{\textbf{x}}_k^i)| \phi^{i}_k \rangle }{\langle \Psi_T|\phi^{i}_k \rangle }\,.
\end{aligned}
\label{eq:importance_function}
\end{equation}
The vector shift $\overline{\textbf{x}}^{i}_k$ is chosen to minimize the fluctuation of the weights and the optimal choice 
for each component
is given, at small $\tau$,
by \cite{AFQMC_shift}:
\begin{equation}
\begin{aligned}
\overline{x}^{i}_{k,\gamma} 
=-\sqrt{\tau}\frac{\langle  \Psi_T|\hat{L}_\gamma |\phi^i_k \rangle}{\langle \Psi_T| \phi^i_k\rangle}\,.
\end{aligned}
\label{eq:shift1}
\end{equation}
With walkers propagated by the above procedure 
to the $n$-{th} step that  
reach equilibrium (i.e., with $\beta_{\rm eq}\equiv n_{\rm eq}\tau$ large enough to  
reach the ground state in Eq.~(\ref{eq:imaginary_time_evolve}) within the targeted statistical accuracy),
the measurement of ground-state energy can be taken by the mixed estimator~\cite{AFQMC_Zhang-Krakauer-2003-PRL,WIREs_Mario} for all $n\ge n_{\rm eq}$ 
\begin{equation}
\frac{\langle \Psi_T |\hat{H} |\Psi^n \rangle}{\langle \Psi_T |\Psi^n \rangle}=\frac{\sum_{k} W^{n}_k\frac{\langle \Psi_T |\hat{H} |\phi^{n}_k\rangle}{\langle \Psi_T|\phi^{n}_k\rangle}}{\sum_{k } W^{n}_k}.
\label{eq:mixed_estimator}
\end{equation}


The importance sampling formalism above together with the force bias automatically imposes the constrained path approximation~\cite{QMC_Zhang_Constrained_1997,lecturenotes-2019} in the presence of a sign problem. However, when there is a phase problem, e.g., with general electronic Hamiltonians as 
in Eq.~(\ref{eq:H_MC}) 
and thus complex auxiliary fields as in Eq.~(\ref{eq:HS-V}), an 
additional ingredient is necessary to complete the phase-free or phaseless approximation~\cite{AFQMC_Zhang-Krakauer-2003-PRL,lecturenotes-2019,Hao_Some_recent_developments}: 
\begin{equation}
I_{\textup{ph}}(\textbf{x}, \overline{\textbf{x}}^{i}_k, \phi^{i}_k) = \bigl|I(\textbf{x}, \overline{\textbf{x}}^{i}_k, \phi^{i}_k)\bigr|\,* \textup{max}(0, \textup{cos}(\Delta \theta))\\
\label{eq:importance_function_CP}
\end{equation}
with 
\begin{equation}
\Delta\theta=\textup{Arg}\frac{\langle \Psi_T| \phi^{i+1}_k \rangle }{\langle \Psi_T|\phi^{i}_k \rangle }.
\end{equation}
The implementation details can vary depending on the approach (e.g., local energy vs. hybrid methods \cite{hybrid_phaseless}, removing the absolute value 
in Eq.~(\ref{eq:importance_function_CP})
by preserving the overall phase during long projection times \cite{AFQMC_Zhang-Krakauer-2003-PRL, AFQMC_PathRestoration}, or using cosine projection vs. the half-plane method or other techniques to eliminate the finite density at the origin in the complex plane \cite{Hao_Some_recent_developments}). In cases where all auxiliary fields are real (such as for Hubbard interactions), the phaseless formalism reduces to the constrained path approach \cite{QMC_Zhang_Constrained_1997}, $\Delta \theta = 0$, and the zero values in the importance function prevent the random walk from encountering the sign problem. The implementation of constraints introduces a systematic bias that depends on how well the trial wave function $\langle \Psi_T|$ can characterize the ground state. Extensive benchmark studies have demonstrated that AFQMC, under this formalism and theory \cite{QMC_Zhang_Constrained_1997, AFQMC_Zhang-Krakauer-2003-PRL}, achieves high accuracy, both in model systems and real materials \cite{simons_material_2020, Hubbard_benchmark_2015}.

\subsection{\label{sec:Prilimilaries_wf} Explicitly correlated trial wave functions $|\Psi_T\rangle$}

In this section we describe the form of the correlated wave functions considered in this work.
We will be concrete and explicit in  defining 
the wave functions. This is not meant to imply that the algorithm 
is limited to the specific 
form, but rather to 
make the ensuing technical discussions more easily accessible.

A large class of explicitly correlated wave functions can be 
represented by the general form
\begin{equation}
|\Psi_T \rangle  = \prod_{j=1}^m e^{-t_j \hat {\mathcal H}_T^j } | \Psi_T^0\rangle\,, 
\label{eq:trial}
\end{equation}
where the operator $\hat {\mathcal H}_T^j$ can be thought of 
as {\it effective\/} Hamiltonians. For example, if $| \Psi_T^0\rangle$ is chosen 
as a mean-field solution, a Slater determinant, $m=1$, and $\hat {\mathcal H}_T^1$ is a 
density-density two-body operator 
(e.g., following a variational optimization), then $|\Psi_T \rangle $ is a Slater-Jastrow wave function. 
Somewhat more generally but still keeping $m=1$, if we choose  $| \Psi_T^0\rangle$ to be the Hartree-Fock (HF) solution, and set the operator 
$-t_1 \hat {\mathcal H}_T^j = \hat T_1 +\hat T_2$, where $\hat T_1$ and $\hat T_2$ are 
the corresponding single- and double-excitation operators respectively, we recover
the CCSD wave function.
The general form in Eq.~(\ref{eq:trial}) also represents the wave function in variational AFQMC (VAFMQC)~\cite{VAFQMC_Sorella,HAFQMC,VAFQMC_Ryan}, where $\hat {\mathcal H}_T^j$ is an effective 
two-body Hamiltonian whose one-body and two-body parts are variationally optimized,
as further summarized in the Appendix.~\ref{sec:APPENDIX_VAFQMC}. 

As discussed at the end of Sec.~\ref{sec:Prilimilaries_overview}, all of these wave function forms can be cast in the form
\begin{equation}
|\Psi_T \rangle  = 
\prod_{j=1}^{m} \int  p_T(\textbf{y}^j)
\hat {\mathcal B}_T^j (\textbf{y}^j) d\textbf{y}^{j}\, | \Psi_T^0\rangle\,, 
\label{eq:trial-AF}
\end{equation}
where, as in Eqs.~(\ref{eq:HS-V}) and (\ref{eq:HS}), $\textbf{y}^{j}$ are auxiliary-fields 
from decoupling any two-body terms in $\hat {\mathcal H}_T^j $, and we have combined successive products and 
used  $\hat {\mathcal B}$ to denote exponentials of one-body operators.
(We use $\hat {\mathcal B}$ for trial wave function and  $\hat B$ for those
in AFQMC, which come from the actual Hamiltonian $\hat H$.)
When convenient we will use the following 
more compact form to represent the above:
\begin{equation}
|\Psi_T \rangle  =  \int  p_T(Y) \hat {\mathcal B}_T (Y) dY\, | \Psi_T^0\rangle\,, 
\label{eq:trial_decomp}
\end{equation}
where 
$Y=\{\textbf{y}^j\}$.
Note that, as in the initial wave function of AFQMC, it is straightforward 
to incorporate a multi-reference initial 
wave function $| \Psi_T^0\rangle$ in Eq.~(\ref{eq:trial_decomp}), by for example directly sampling a linear combination of Slater determinants \cite{Hao_Some_recent_developments}.

\subsection{\label{sec:Prilimilaries_sampling} BRW and MCMC sampling in AFQMC}

In this section, we summarize the two kinds of sampling algorithms which have been used 
for auxiliary-field calculations, namely MCMC with the generalized Metropolis algorithm, and BRW with a population of walkers in branching random walks. The purpose is 
to help illustrate their connection and difference, and establish a framework for describing the new algorithm which uses a combination of the two. 

For concreteness, let us consider the evaluation of the following 
\begin{widetext}
\begin{equation}
\begin{aligned}
\langle \hat{H} \rangle =\frac{\langle \Psi_T| \hat{H} |\Psi_{0}\rangle}{\langle \Psi_T|\Psi_{0}\rangle}&=\frac{\int \!  \prod_{j=1}^{m}\!\! d\textbf{y}^{j}\!  \prod_{i=1}^{n}\!\! d\textbf{x}^{i} \langle \phi_T|\prod_{j=1}^{m}\!\! p_T(\textbf{y}^{j})\hat{\mathcal B}_T^j(\textbf{y}^{j})\,\hat{H}\,\prod_{i=1}^{n}\!\!p(\textbf{x}^{i})\hat{B}(\textbf{x}^{i})|\phi_I\rangle}{\int \! \prod_{j=1}^{m}\!\! d\textbf{y}^{j}\! \prod_{i=1}^{n}\!\!d\textbf{x}^{i}\,\langle  \phi_T| \prod_{i=1}^{m}\!\!p_T(\textbf{y}^{j})\hat{\mathcal B}_T^j(\textbf{y}^{j})\,\prod_{i=1}^{n}\!\!p(\textbf{x}^{i})\hat{B}(\textbf{x}^{i})|\phi_I\rangle }\,, 
\label{eq:overview2}
\end{aligned}
\end{equation}
\end{widetext} 
where 
we have used the notations introduced in Secs.~\ref{sec:Prilimilaries_overview} and \ref{sec:Prilimilaries_wf} and, for simplicity, chosen 
 $| \Psi_I\rangle$ in Eq.~(\ref{eq:imaginary_time_evolve}) and $| \Psi_T^0\rangle$ 
 in Eq.~(\ref{eq:trial}) to both be single Slater determinant wave functions, denoted by
$| \phi_I\rangle$ and 
$| \phi_T\rangle$, respectively. (It should be understood here and below that 
a $^\dagger$ is implied as needed when
a propagator, such as $\hat{\mathcal B}_T^j$ above,
is non-hermitian and is written in bra form.)  

\begin{figure}[htbp]
\includegraphics[scale = 0.25]{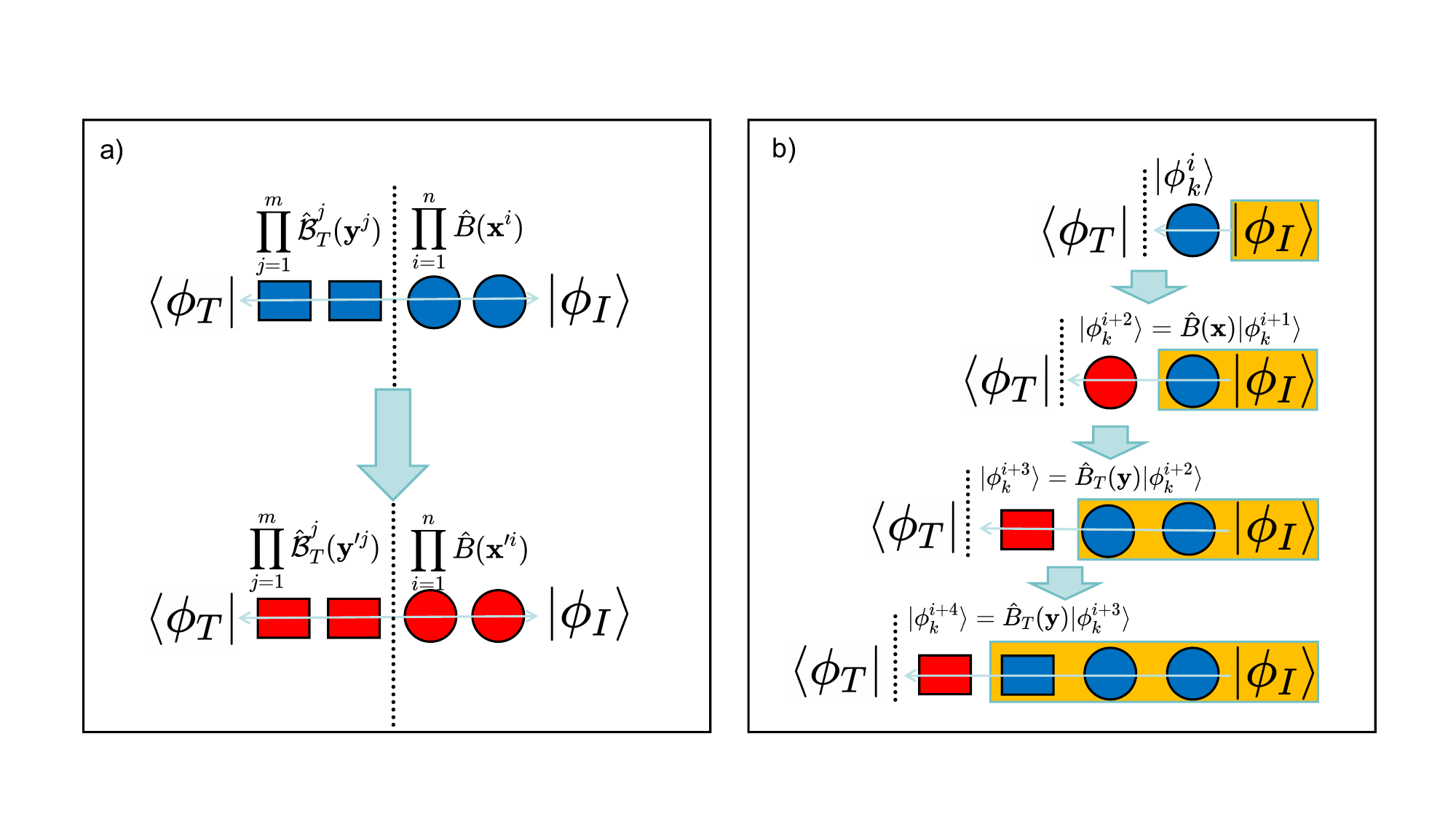} 
\caption{Illustration of 
(a) the MCMC and (b) the BRW algorithms to evaluate Eq.~\ref{eq:overview2}. 
The figures depict $m=2$ and $n=2$. Circle and rectangular denote one-body operators $\hat B(\textbf{x})$ and $\hat {\mathcal B}_T^j(\textbf{y})$, respectively. 
Blue/red indicates current/updated state in the MC. 
In (a), paths of auxiliary-fields,  $\{\textbf{y}^j,\textbf{x}^i\}$, 
and sampled by the Metropolis algorithm. In the figure, a schematic is shown 
 for one path. 
In (b), a population of walkers $\{|\phi^i_k\rangle\}$ (with weight) are propagated from right to left.  In the figure, the trajectory of one walker (labeled by $k$) is shown. 
The rectangular box shaded orange grows from right to left, indicating that often in ground-state 
calculations only the current walker state is kept and the 
AF path history can be discarded.
}
\label{Fig.Metro_BRW} 
\end{figure}

The two flavors of Monte Carlo to 
evaluate Eq.~(\ref{eq:overview2}),
from a high-level algorithmic standpoint, 
can be summarized as
\begin{itemize}
    \item Metropolis or Markov chain Monte Carlo (MCMC), illustrated in Fig.~\ref{Fig.Metro_BRW}(a). In this approach, the Monte Carlo sampling is performed on a ``chain" of auxiliary-fields 
    $(Y,X)$, where 
    $X\equiv \{ \textbf{x}^{n} \cdots,\textbf{x}^{1}\}$.
 This chain has fixed length $m+n$, moves with updates proposed 
 $(Y,X)\rightarrow (Y',X')$, 
 and accepted/rejected by the generalized Metropolis algorithm.
Typically, one sweeps through the chain (e.g., from right to left and then left to right) for these moves.
   
    \item BRW, illustrated in Fig.~\ref{Fig.Metro_BRW}(b). In this approach, the Monte Carlo sampling typically carries a population indexed by $k$, and the branching random walk proceeds from right to left to advance the random walker via the sequential sampling of the AFs: $\{|\phi^i_k\rangle,W^i_k\} \xrightarrow{\textbf{x}^{i}_k} 
 \{|\phi^{i+1}_k\rangle,W^{i+1}_k\} 
 \cdots    $.
\end{itemize}

We make a few remarks to help connect the above to broader contexts. 
If we take $\hat B=\hat {\mathcal B}_T^j$ and $\phi_I=\phi_T$, then 
Fig.~\ref{Fig.Metro_BRW}(a) depicts the sampling in VAFQMC~\cite{HAFQMC,VAFQMC_Ryan}.
On the other hand, 
if we take  $\hat {\mathcal B}_T^j=\hat B$, then Fig.~\ref{Fig.Metro_BRW}(a) illustrates 
the usual path-integral or projector AFQMC formulation. Under this  
choice, Fig.~\ref{Fig.Metro_BRW}(b) depicts the AFQMC algorithm in Sec.~\ref{sec:Prilimilaries_AFQMC}, with $\langle \phi_T|$ as trial wave function. In the latter case, the squares can be viewed as back-propagation \cite{AFQMC_PathRestoration}.
In both cases, any observable $\hat O$ can be inserted in place of $\hat H$ to compute the expectation $\langle O\rangle$. As mentioned, the BRW approach~\cite{QMC_Zhang_Constrained_1997, AFQMC_Zhang-Krakauer-2003-PRL, lecturenotes-2019} is necessary because
 it is otherwise difficult to impose a constraint without incurring a global ergodicity problem. 
In Ref.~\cite{zxiao_MRC_2023}, a method was introduced to interface 
BRWs with Metropolis sampling to realize constraint release. That approach can 
be understood with the help of Fig.~\ref{Fig.Metro_BRW}(b). The BRW sampling
is first performed,
including the ``square" portion, as if in an AFQMC calculation with back-propagation. This leaves 
us with $\{|\phi^n_k\rangle, W_k^n \}$ and paths $\{Y_k\}$.
We then sample the paths $\{Y_k\}$ by Metropolis with respect to the probability function
$\big|\langle \phi_T|  p(Y_k)
\hat B (Y_k) | \phi_k^n\rangle\big|$. 
Inclusion of the sign/phase factor 
in computation of $\langle \hat O\rangle$ (inserted next to $| \phi_k^n\rangle$) achieves 
constraint release \cite{zxiao_MRC_2023}.

The use of a trial wave function which is itself written as an integral over 
auxiliary or hidden variables, however, requires a significant, additional step 
in the method. Naive sampling of the trial wave function would lead to loss of fidelity and large bias. This is the focus of the present work, and the rest of the discussion.

\section{\label{sec:AFQMC_Metro} Constraint and importance sampling by MCMC}

\subsection{\label{sec:results_method}Formalism}

As discussed in Sec.~\ref{sec:Prilimilaries_AFQMC},
the trial wave functipon $|\Psi_T\rangle$ appears in
the realization of BRW sampling and the 
implementation of constraints in AFQMC 
only in the ratio of
overlaps 
and ``local'' quantities 
(force bias and local energy).
This is a key 
that makes it possible to devise an efficient and reliable implementation of $|\Psi_T\rangle$ by stochastic sampling.

\begin{figure}[htbp]
\includegraphics[scale = 0.3]{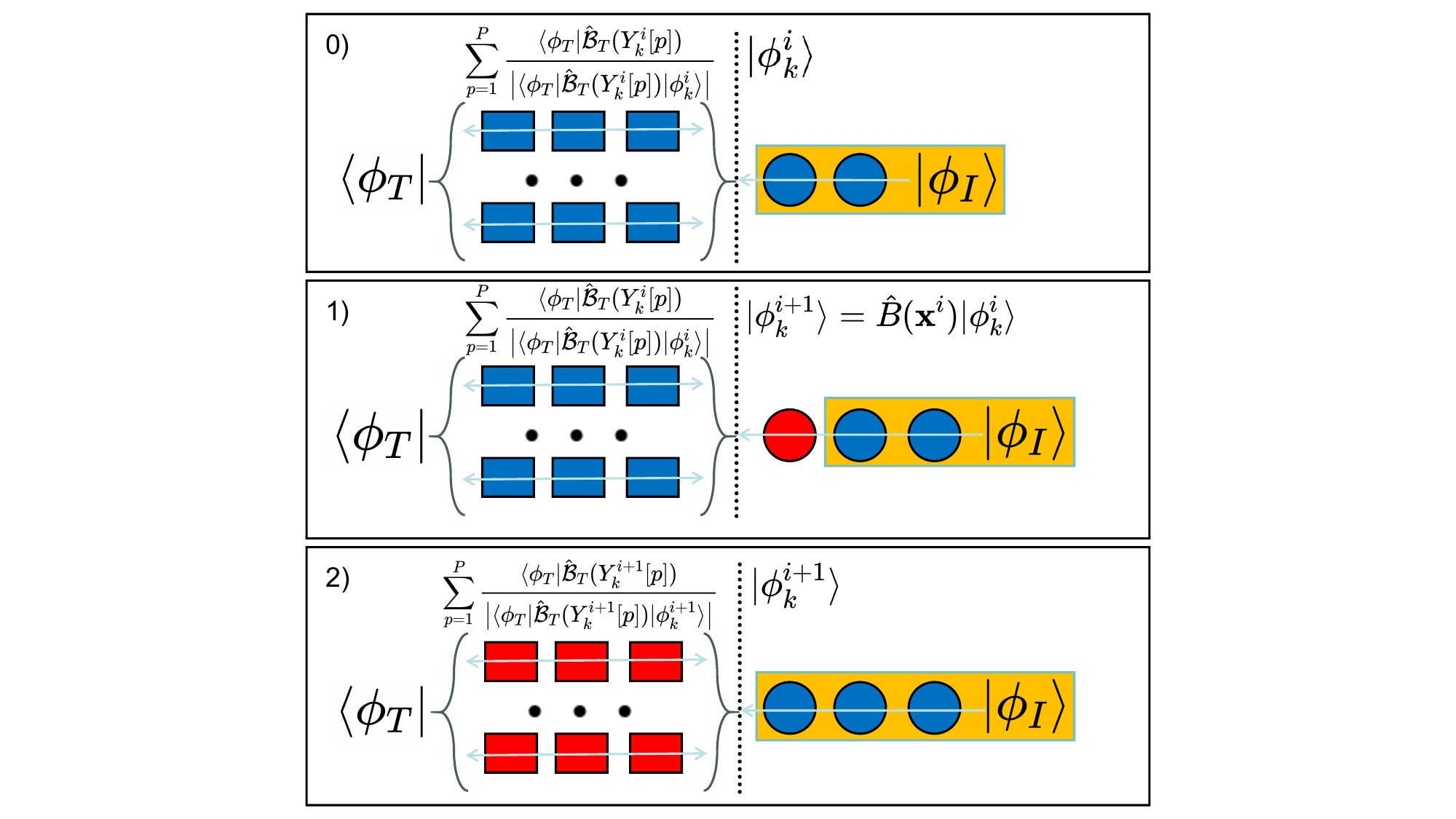} 
\caption{
Illustration of the structure of our Monte Carlo sampling algorithm. 
One random 
walker and its tethered MCMC paths are shown, during one leapfrog update step.
The colors blue/red/orange have the same meaning as
in Fig.~\ref{Fig.Metro_BRW}, and circle and rectangular again denote $\hat B$ and $\hat {\mathcal B}_T$, respectively.
(0) current state of the walker $|\phi^{i}_k \rangle$ and the $|\Psi_T \rangle$ paths attached to it. (Taken literally, this picture depicts $i=2$, with $m=3$ and $P$ given by the number of rows on the left.)
(1). the walker is propagated one step, $|\phi^i_k \rangle \rightarrow |\phi^{i+1}_k \rangle$ and its weight updated $W^i_k  \rightarrow W^{i+1}_k$,
using the current samples of  $\langle \Psi_T|$
as importance function and constraint. 
(2). MCMC sweeps are performed to update each path (each $p$, represented by one row of squares), 
$Y_k^i[p] \rightarrow Y_k^{i+1}[p]$, according to the walker's new position, $|\phi^{i+1}_k \rangle$. 
}
\label{Fig.AFQMC_Metro} 
\end{figure}

Our algorithm consists of a population of random walkers $\{|\phi^{i}_k\rangle, W_k^i\}$ (as in standard AFQMC), but additionally with each walker tethered to a set of $P$
paths $\{Y_k^i[p]\}$, with $p=1,2,\cdots, P$. 
The walker and its paths are moved stochastically in a leapfrog pattern, 
with the walker by BRW and the paths by MCMC.
The structure of this ``composite walker'', the random walker plus its connected paths,
and how the leapfrog move is executed, is illustrated in 
Fig.~\ref{Fig.AFQMC_Metro}.
We next discuss the leapfrog sampling in detail, 
by focusing 
on one walker, labeled $k$, and one step from $i$ to $i+1$. To simplify the notation below, we will
omit these indices, using $|\phi\rangle$, $W$, and $Y[p]$ to describe the corresponding objects at $i$, and primed quantities  $|\phi'\rangle$, $W'$, and $Y'[p]$ for $i+1$,
i.e., unprimed for the blue objects and primed for red in Fig.~\ref{Fig.AFQMC_Metro}.

Step 1 of the leapfrog move is 
$|\phi\rangle \rightarrow |\phi'\rangle$.
Suppose
we have $P$ paths  $\{Y[p]\}$ which 
are sampled from the probability density function
\begin{equation}
{\mathcal P}(Y; \phi) = 
\big|\langle \Psi_T^0| p_T(Y) \hat {\mathcal B}_T(Y) |\phi\rangle\big|/{\mathcal N}(\phi)\,,
\label{eq:PDF-Y}
\end{equation}
with normalization 
${\mathcal N}(\phi)\equiv
{\int \big|\langle \Psi_T^0| p_T(Y) \hat {\mathcal B}_T(Y) |\phi\rangle\big|\,dY}$
(which is not known or needed).
Then we have a Monte Carlo estimate of the $\langle \Psi_T|$ given in Eq.~(\ref{eq:trial_decomp}):
\begin{equation}
\langle \Psi_{T}| \doteq \langle \bar \Psi_{T}|
=\frac{{\mathcal N}(\phi)}{P} \sum_{p=1}^P 
\frac{\langle \phi_T| 
\hat{\mathcal B}_T(Y[p]) }
{\big| \langle \phi_T| 
\hat{\mathcal B}_T(Y[p])|\phi\rangle\big|}\,.
\label{eq:Metropolis_sampled_trial}
\end{equation}
Here and below we use the ``bar'' as an explicit indication of particular 
Monte Carlo estimate of $\langle \Psi_T|$.
(For simplicity we will consider the case where $\langle \Psi_T^0|$ is a single Slater determinant, hence the replacement by $\langle \phi_T|$ in Eq.~(\ref{eq:Metropolis_sampled_trial}) and below. The discussion below can be generalized straightforwardly to  the case of a linear combination of Slater deterimnants following the usual procedures.) 
The overlap ratio can then be estimated by: 
\begin{equation}
\frac{\langle \bar \Psi_{T}|\phi' \rangle}{\langle \bar \Psi_{T}|\phi \rangle}
=
\frac{\sum_{p=1}^P 
\frac{\langle \phi_T| 
\hat{\mathcal B}_T(Y[p])|\phi'\rangle}{\langle \phi_T| 
\hat{\mathcal B}_T(Y[p])|\phi\rangle}\,S(Y[p])}
{\sum_{p=1}^P S(Y[p])}\,,
\label{eq:ratio-update-phi}
\end{equation}
where 
\begin{equation}
S(Y[p])\equiv  
\frac{\langle \phi_T| 
\hat{\mathcal B}_T(Y[p])|\phi\rangle}{\Big|\langle \phi_T| 
\hat{\mathcal B}_T(Y[p])|\phi\rangle\Big|}
\label{eq:sign-path}
\end{equation}
is a phase factor for each of the  $P$ paths associated with the walker $|\phi\rangle$. Any local expectation of $|\phi\rangle$ can also be 
estimated as 
\begin{equation}
\frac{\langle \bar \Psi_{T}|\hat O |\phi \rangle}{\langle \bar \Psi_{T}|\phi \rangle}\doteq
\frac{\sum_{p=1}^P 
\frac{\langle \phi_T| 
\hat{\mathcal B}_T(Y[p])\,\hat O\,|\phi\rangle}{\langle \phi_T| 
\hat{\mathcal B}_T(Y[p])|\phi\rangle}\,S(Y[p])}
{\sum_{p=1}^P S(Y[p])}\,.
\label{eq:local-obs}
\end{equation}
This allows us to compute the force bias in Eq.~(\ref{eq:shift1}), with which the propagation in Eq.~(\ref{eq:walker_prop}) 
can be completed. 
The weight update $W\rightarrow W'$
is accomplished through  
the function 
$I_{\rm ph}$ in Eq.~(\ref{eq:importance_function_CP}).
In the hybrid formalism of ph-AFQMC, 
$I$ as written in Eq.~(\ref{eq:importance_function})
can be computed using the overlap ratio formula in Eq.~(\ref {eq:ratio-update-phi}).
In the local energy formalism, $I$ is given by $e^{-\tau E_L(\phi)}$, in which 
the local energy $E_L(\phi)$ can be computed by Eq.~(\ref{eq:local-obs}).
In either case, $\Delta \theta$ is given by the overlap ratio. 

Step 2 of the leapfrog move is to update the paths $\{Y[p]\}\rightarrow \{Y'[p]\}$,
conditional on the new walker position $|\phi'\rangle$ which has now been obtained following step 1. 
The goal is to generate the new set of paths,
$\{Y'[p]\}$,
distributed according to ${\mathcal P}(Y';\phi')$, as defined in Eq.~(\ref{eq:PDF-Y}).
For each of the $P$ paths, denoted by $Y'$,
we start from the initial state 
inherited from $|\phi\rangle$.
We use the generalized Metropolis algorithm to sample a new $Y'$ by MCMC for $M$ steps. In  
each step a new state $Y'_{\rm prop}$ is proposed. 
For example, if no force bias is used, one could propose from the probability density function $p_T(Y'_{\rm prop})$, in which case the  probability 
to accept $Y'_{\rm prop}$ as the new $Y'$ 
is given by $\min\{1, \big|\langle \phi_T|\hat {\mathcal B}_T(Y'_{\rm prop}) |\phi'\rangle\big|/\big|\langle \phi_T|\hat {\mathcal B}_T(Y') |\phi'\rangle\big|\}$.
Alternatively, force bias or hybrid Monte Carlo, 
or any other sampling schemes one might use in VAFQMC, can be applied. 
Since $|\phi\rangle$ is close to $|\phi'\rangle$, 
to re-equilibrate or re-thermalize the paths from 
${\mathcal P}(Y';\phi)$ to ${\mathcal P}(Y';\phi')$ only requires a small $M$. 
At the end of the MCMC, the paths $ \{Y'[p]\}$ sample ${\mathcal P}(Y';\phi')$.


In the above (leapfrog step 2), 
a ``hand-off'' occurs between two stochastic samples of the trial wave function, $|\bar \Psi_T'\rangle $ and $|\bar \Psi_T\rangle $. Viewed in terms of the 
importance sampling transformation, an extra ratio of overlaps is inserted 
prior to the beginning of the next AFQMC step: ($|\phi'\rangle \rightarrow |\phi''\rangle$):
\begin{equation}
R(\bar \Psi_T\to \bar\Psi_T';\phi')=\frac{\langle \bar \Psi_T'| \phi' \rangle }{\langle \bar \Psi_T|\phi'\rangle }\,.
\label{eq:AFQMC_Metro_correction}
\end{equation}
The ratio in Eq.~(\ref{eq:AFQMC_Metro_correction}) should of course be unity when 
evaluated explicitly. However, the statistical sampling of $\langle \Psi_T|$ introduces noise, including a possible phase factor. If the overlap $\langle \Psi_T|\phi'\rangle$ is small, there can even be a sign change. 
To suppress this noise and prevent it from destabilizing the calculation, we
apply a step following the philosophy of the phaseless constraint. We
modify the weight of the walker $|\phi'\rangle$ by the cosine projection:
\begin{equation}
W'\rightarrow W'\,\cdot \max(0,\cos(\Delta\theta_T))\,,
\label{eq:AFQMC_Metro_constraint_cos}
\end{equation}
where $\Delta\theta_T=\textup{Arg}[R(\bar \Psi_T\to \bar\Psi_T';\phi')]$.
Alternatively, we can apply the ``half-plane'' version~\cite{AFQMC_Zhang-Krakauer-2003-PRL}, i.e.,
if $\abs{\Delta\theta_T}>\pi/2$,  the walker 
$|\phi' \rangle $ is eliminated ($W'=0$). The two approaches did not show noticeable difference in our tests, but the former yielded somewhat smaller statistical uncertainties. One could combine Eq.~(\ref{eq:AFQMC_Metro_constraint_cos}) with the AFQMC projection with $\Delta\theta$ in leapfrog step 1, by adding the two angles and performing one 
``cosine projection", which is equivalent in the small angle limit (sufficiently 
small $\tau$ and large enough $P$). 
Most of the data presented below are obtained using this variant.
We comment that yet another way to 
suppress the influence of the noise can be 
the penalty method in Ref.~\cite{Ceperley_penalty_MCMC}, which should lead to similar results. 

A key to the success of sampling $|\Psi_T\rangle$ is 
the severity of the phase or sign fluctuations in the trial wave function. 
The mean of $S(Y[p])$ in Eq.~(\ref{eq:sign-path}) will determine the variance 
in our calculation and how severe 
the noise will be, including the phase fluctuation of $\Delta\theta_T$. 
 The precursor for this can be found in the variational calculation using $|\Psi_T\rangle$. If the 
sign/phase problem is severe there, the ensuing 
AFQMC calculation will be more noisy and will require more samples (e.g., larger $P$).

As a technical remark, the structure of the composite walker provides considerable 
flexibility for optimizing computational efficiency. For example, 
 the mixed estimator for the 
energy at step $i$ is normally computed 
with the trial wave function $\langle \bar \Psi^i_{T,k}|$, using the $P$ paths tethered to the walker $|\phi^i_k\rangle$. However, if a higher accuracy is 
desired in the measurement, we could sample additional paths with more MCMC sweeps, to obtain 
$\langle \bar \Psi'^i_{T,k}|$ to  evaluate the local energy. All that is needed is an additional correction factor $R(\bar \Psi^i_{T,k}\to \bar \Psi'^i_{T,k};\phi^i_k)$ in the numerator and denominator in Eq.~(\ref{eq:mixed_estimator}).

In the APPENDIX.~\ref{sec:APPENDIX_summary}, we include a detailed outline of our algorithm in the form of a pseudo-code. 

\subsection{\label{sec:results_F}Illustrative example}

Before discussing realistic applications and presenting results, 
we first study a toy problem to help illustrate some of the characteristics and behavior
of the new method. 
In this test, we consider a 
more straightforward choice of the 
 trial wave function given in the general form of  
Eq.~(\ref{eq:trial}), by 
letting $\hat {\mathcal H}_T^j=\hat H$ and $t_j=\tau$
(with $j=1,\dots, m$), 
and taking $|\Psi_T^0\rangle=|\phi_{\rm RHF}\rangle$, the 
restricted Hartree-Fock (RHF) solution.
This leads to a trial wave function 
$|\Psi_T\rangle=e^{-\beta_T\hat H}|\phi_{\rm RHF}\rangle$
with $\beta_T=m\tau$, which
can be viewed as the wave function produced by the same AFQMC projection, without constraint. As $m$ is increased,  $|\Psi_T\rangle$ is improved by a longer projection, and eventually approaches the exact ground state. This allows us to study the convergence of our method systematically by varying $\beta_T$.

\begin{figure}[htbp]
\includegraphics[scale = 0.2]{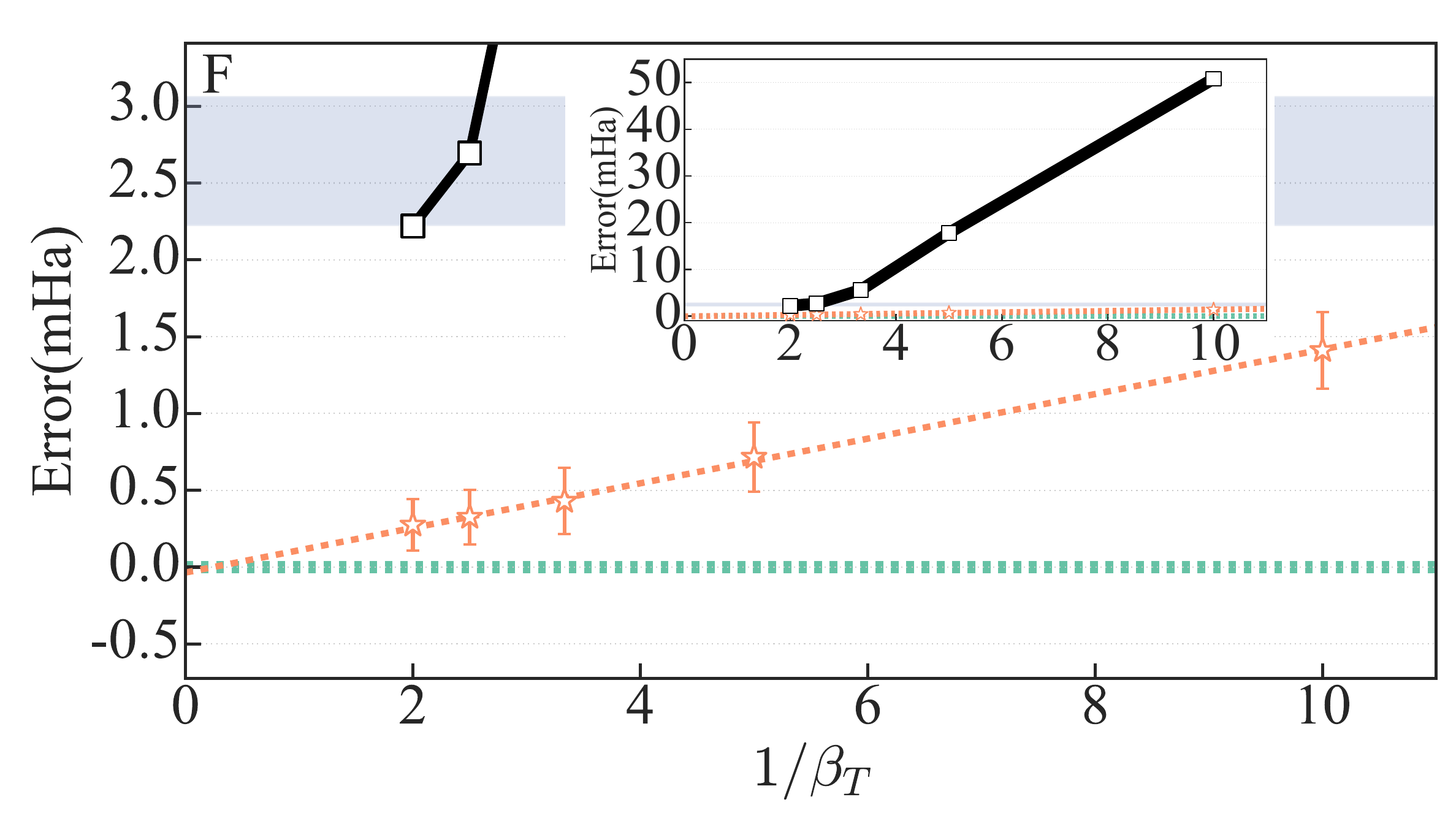} 
\caption{Ground state energy 
systematic errors for AFQMC with different trial wave functions on the F atom.
The trial wave function is $|\Psi_T\rangle=e^{-m\,\tau\,\hat H}|\phi_{\rm RHF}\rangle$ with $\tau = 0.01$.
The cc-pVDZ basis
is used.
The error is given by $E_{\rm AFQMC}-E_{\rm FCI}$ and $\beta_T = m\tau$ is also given in atomic units. 
The AFQMC systematic error with the RHF itself as trial wave function is presented as the ``shaded blue background" whose width indicates the statistical errorbar. ``Orange star" denotes AFQMC results, and related trial energy is specified with ``black square".
} 
\label{Fig.F} 
\end{figure}

The result is presented
in Fig.~\ref{Fig.F} for the F atom, where we show the computed ground state energy 
vs.~the inverse of the projection time $\beta_T$ in $|\Psi_T\rangle$. 
With $m=0$, we have $\beta_T=0$, and our algorithm reduces to standard 
phaseless AFQMC with the RHF trial wave function.
A constraint bias of $\sim 2.5$\,mHa is observed, which reflects the deficiency of the RHF trial wave function for open-shell systems and is consistent with previous results~\cite{wirawan-F2-spin-filtration}.
 As $\beta_T$ is increased with increasing $m$, the systematic error is reduced accordingly. With $\beta_T\sim 0.2$, the computed total energy is already within chemical accuracy of the exact result
 (obtained from full configuration interaction (FCI) with PySCF \cite{pyscf}).
The trial wave function is further improved 
as  $\beta_T$ is further increased, and approaches the exact result.

Although the above ``imaginary-time evolved" $ |\Psi_T\rangle$ is useful for a systematic study  in simple systems, it also highlights the main challenges 
in producing correlated  trial wave functions which scale properly with system size.
To achieve consistent results in larger systems, the $\beta_T$ required would need to increase. An additional technical difficulty is that the use of the ``physical''
Hamiltonian $\hat H$ and ``realistic'' time-step are both non-optimal for producing a good trial wave function. Many time-slices are involved and typically a significant phase problem arises even for relatively short projection time (same as in the free projection method \cite{AFQMC_Zhang-Krakauer-2003-PRL}). 
The general form in Eq.~(\ref{eq:trial}) allows better choices, one of which is variational AFQMC  \cite{VAFQMC_Sorella, VAFQMC_Ryan, HAFQMC}. Below we will 
use the particular of variational wave function from Ref.~\cite{HAFQMC} as our trial wave function.

\section{\label{sec:Results}Results}

In this section, we test the algorithm on more strongly correlated molecules, where the error from AFQMC with single Slater determinent trial wave functions is typically more pronounced. We consider two representative situations for static and dynamic correlations.
In Sec.~\ref{sec:results_N2} we  study  bond breaking in the N$_2$ molecule, and then in Sec.~\ref{sec:results_CuO_FeO} we study five transition metal oxide molecules
TiO, VO,  CrO, FeO, and CuO. 

We perform
VAFQMC 
to provide trial wave functions for our AFQMC. 
Here, we directly utilize the same VAFQMC algorithm as described in previous Hybrid AFQMC work \cite{HAFQMC} (but without any additional ``free projection" as was often done there). A brief introduction to this algorithm can be found in APPENDIX.~\ref{sec:APPENDIX_VAFQMC}. 
We use the simplest form of $|\Psi_T\rangle$ in the rest of this work, with only a single ``time slice: $m=1$ in  Eq.~(\ref{eq:trial}), and will refer to the resulting $|\Psi_T\rangle$ by VAFQMC throughtout.
More explicitly, 
\begin{equation}
\langle\Psi_{T}|=\langle \phi_T 
|e^{-t_{0} \hat{T}_{0}} \int\prod_{\gamma} \frac{\mathrm{~d} x_{\gamma }}{\sqrt{2 \pi}} e^{-\frac{1}{2} x_{\gamma }^{2}} e^{-t \hat{T}} e^{\sqrt{-s} \sum_{\gamma}^{N_{\gamma}} x_{\gamma } \hat{L}_{\gamma}}\,, 
\label{eq:VAFQMC}
\end{equation}
where all coefficients and $\langle\phi_{T}|$ are optimized to minimize the energy while maintaining a significant sign ratio. 


All VAFQMC and AFQMC calculations here 
are performed in a ``black box" manner (i.e., without tuning parameters for specific systems).
In VAFQMC, the minimum sign ratio is set as $0.7$. The VAFQMC optimization process is iterated for $10000$ steps to ensure the convergence of the energy. 
In our AFQMC, we use a time step $\tau = 0.01$, $2000$ thermal steps, $100 \sim 200$ measurements with 50 steps between each measurement and $240$ walkers for each run. The VAFQMC calculations are carried out on a single NVIDIA A100 GPU. All AFQMC calculations are executed on the “Intel Xeon CPU E5@2.5GHz” CPU platform, with 120 cores for N2 and 240 cores for transition metal oxide molecules.
Additional details specific to the new algorithm are discussed below.
\subsection{\label{sec:results_N2}N2 bond breaking}

Our calculations focus on the N$_2$ molecule at different bond lengths. We use the cc-pVDZ basis, for which there is a wealth of benchmark results from previous studies \cite{AFQMC_bondBreaking, N2_bondbreaking_DMRG}. To present our results systematically, we first illustrate the whole process of our AFQMC/VAFQMC calculations on the ground state energy, and then we elucidate how different parameters in the application of Metropolis constraint can affect our calculations. Finally, we summarize the results for all bond-breaking cases. Our study spans from weakly correlated to strongly correlated regimes, consistently showing highly accurate results with scalable computational costs.  

\begin{figure}[htbp]
\includegraphics[scale = 0.2]{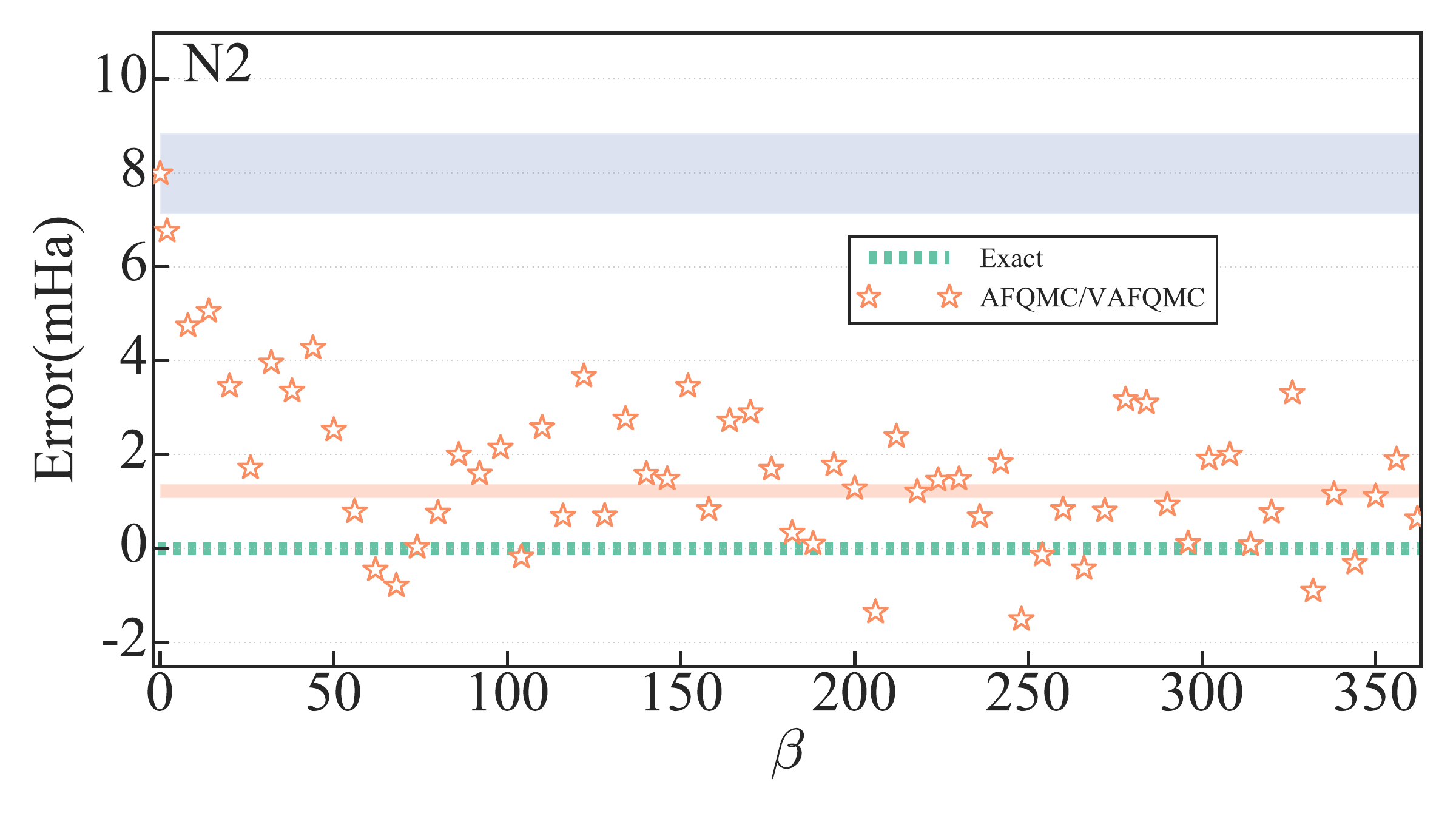} 
\caption{The energy trajectory of the AFQMC propagation with the VAFQMC trial wave function of Eq.~(\ref{eq:VAFQMC}), labeled AFQMC/VAFQMC, in the N$_2$ molecule at a stretched bondlength 
of $R=3.6\,$Bohr. 
The ground-state energy $E(\beta) $ is evaluated 
along the AFQMC imaginary-time propagation with increased $\beta$, and plotted 
as $E(\beta)-E_{\rm DMRG}$, relative
the the near-exact value from DMRG \cite{N2_bondbreaking_DMRG}. 
The converged and final AFQMC ground state energy is shown as 
the shaded orange line whose width indicates the estimated statistical uncertainty. The variational energy of the VAFQMC trial wave function itself is given 
as the shaded blue line, again with the width indicating statistical error.} 
\label{Fig.N2_trajectory} 
\end{figure}

The AFQMC/VAFQMC computational workflow begins 
with the execution of VAFQMC to generate the trial wave function $\langle \Psi_T|$. Subsequently, we chose the initial wave function $|\Psi_I\rangle$ 
in Eq.~(\ref{eq:imaginary_time_evolve})
to be a set of determinants sampled from
$|\Psi_T\rangle$ 
through the Metropolis algorithm. The VAFQMC estimation (i.e., Metropolis sampling) of $\langle \Psi_T|$ then can be seamlessly adopted into the AFQMC initialization. That is, at $\beta=0$, the energy measurement within the AFQMC framework reduces to the Metropolis evaluation of the VAFQMC trial energy. 
We illustrate this by starting the measurement in Eq.~(\ref{eq:mixed_estimator}) at the very beginning of the AFQMC projection, i.e., with
$\beta_{\rm eq}=0$.
The results 
are shown in Fig.~\ref{Fig.N2_trajectory}, 
where the initial AFQMC energy $E(\beta=0)$ is seen to  agree
with the VAFQMC energy shown in the shaded bar.
As $\beta$ increases, the energy decreases and approaches the ground state statistically. For a sufficiently large $\beta$ ($>\beta_{\rm eq}$), the energy $E(\beta)$ will converge, and AFQMC measurements are performed to estimate a final ground state energy together with statistical uncertainty. This AFQMC estimated ground state energy is presented as the shaded orange line. 
The small deviation between AFQMC estimated ground state energy and the exact 
results comes from the constraint bias introduced by the single-time-slice VAFQMC trial wave function in Eq.~(\ref{eq:VAFQMC}). 
This energy deviation of 
$\sim 1.2$mHa is the worst case in our N$_2$ tests.

Our trial wave function $\langle \Psi_T|$ is approximated by the Metropolis samples $\langle \bar \Psi_{T}|$. 
The parameters for the Metropolis sampling, and what statistical accuracy and fidelity it achieves, will 
affect the scaling of the algorithm. 
As introduced in Sec.~\ref{sec:AFQMC_Metro}.A,
these parameters are primarily the number of Metropolis samples, $P$, the number of ``chains" tethered to each walker $|\phi\rangle$,
and the number of thermalization sweeps, $M$ for Metropolis to reach equilibrium. 
To monitor this behavior, calculations are performed with different $P$ and $M$.
Our tests show that the thermalization equilibrium can be reached very quickly. Indeed
often $M=1$ is sufficient; this reflects how related $|\phi^{i+1}_k\rangle$ is to $|\phi^i_k\rangle$, and there is an interplay with the choice of $\tau$.
The number of Metropolis samples that requested to describe $\langle \Psi_T|$ well is also very small. Though our N2 bond-breaking cases ranging from weakly correlated to strongly correlated systems, the required $P$ 
shows little change, and $P=20$ is more than enough for all cases.

\begin{figure}[htbp]
\includegraphics[scale = 0.2]{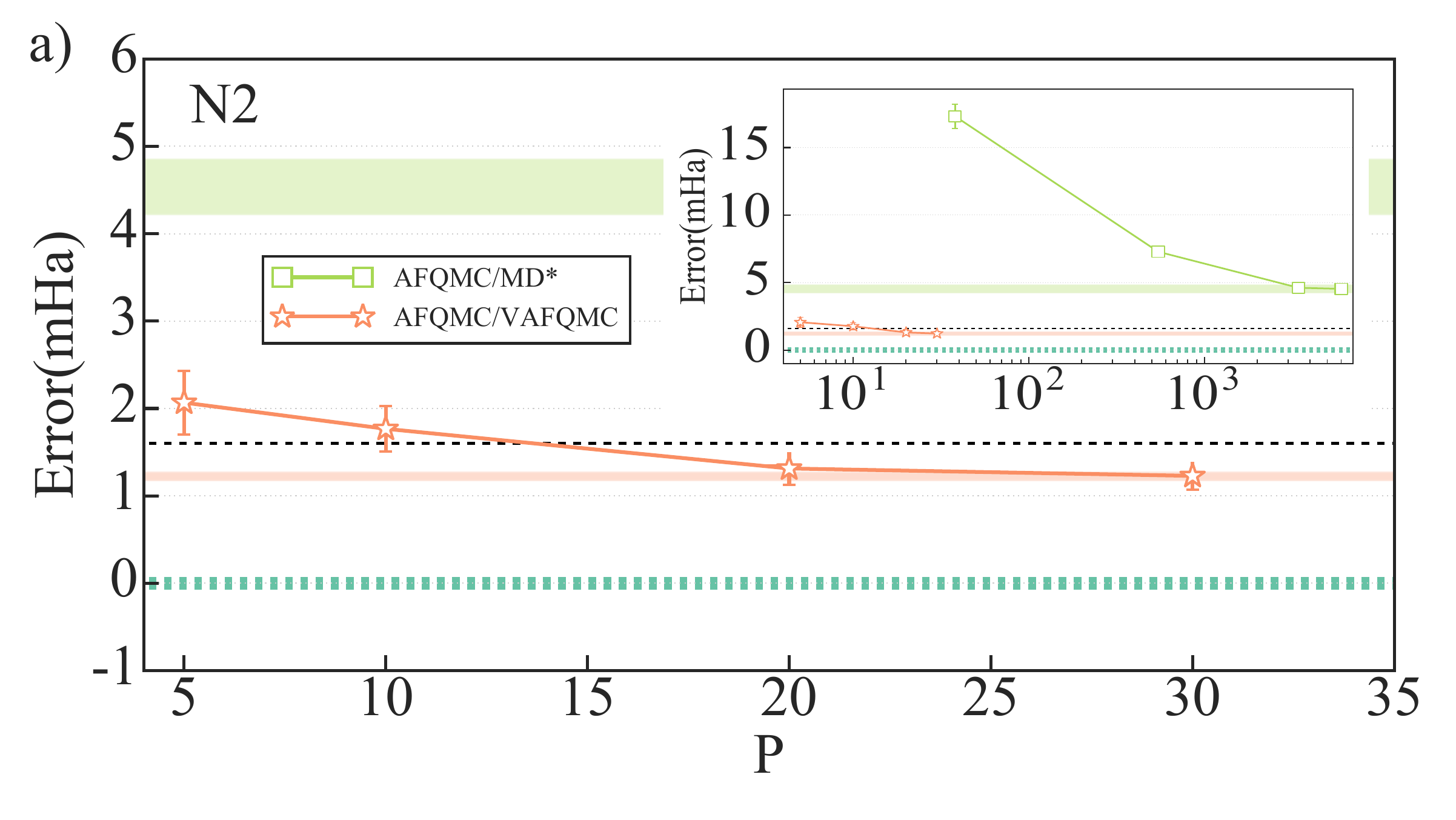} 
\includegraphics[scale = 0.22]{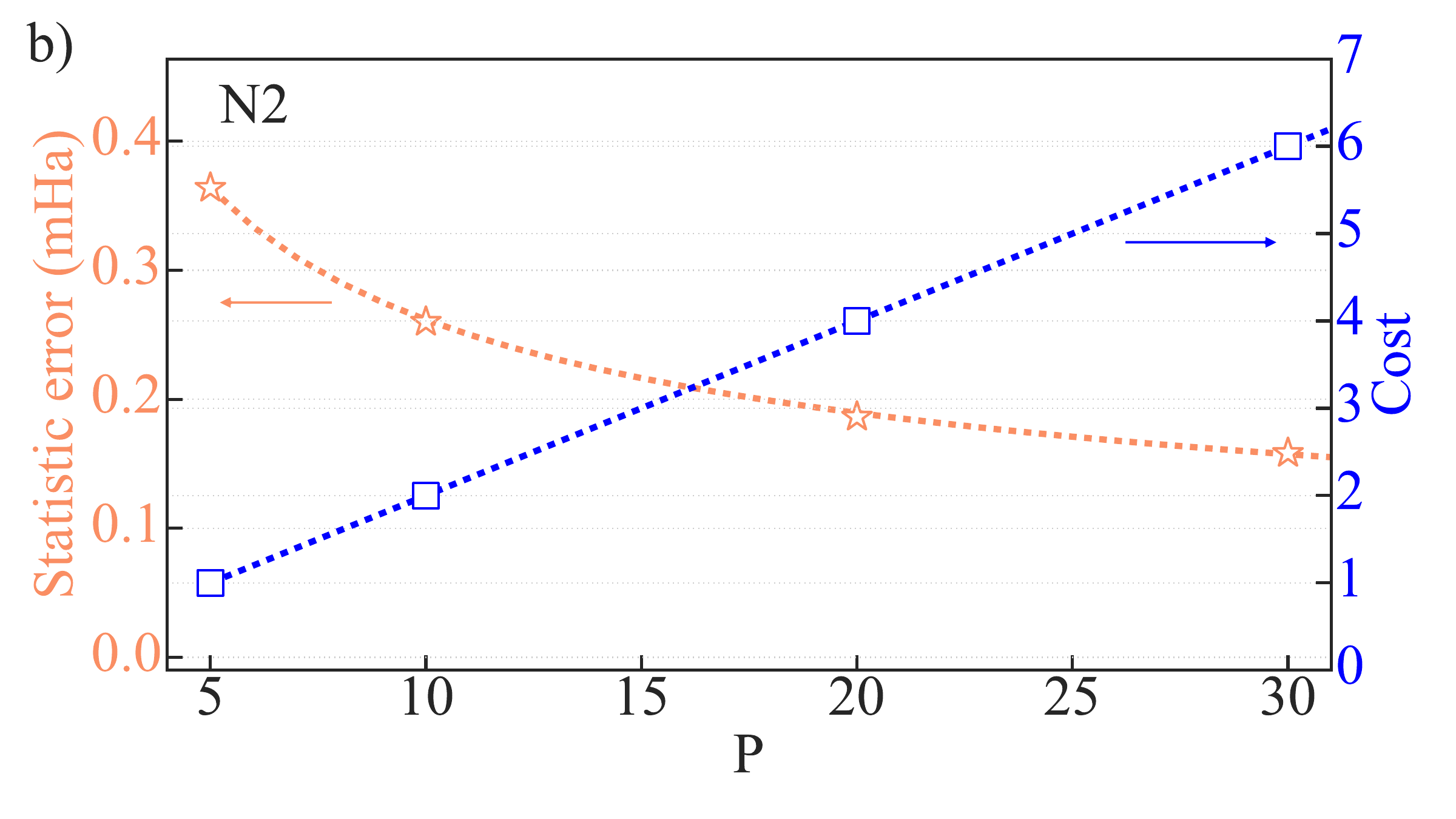} 
\caption{Fidelity and convergence of the stochastic sampling of $|\Psi_T\rangle$, 
and computational efficiency.
(a) Energy computed from AFQMC (relative to the near-exact DMRG results \cite{N2_bondbreaking_DMRG, AFQMC_bondBreaking})
as a function of $P$, the number of paths tethered to each walker in the Metropolis
sampling of the VAFQMC $|\Psi_T\rangle$. The thin dashed black line indicates chemical accuracy. The inset shows, for comparison, results from  AFQMC using a multi-determinant trial wave function taken from CASSCF, truncated to $P$ determinants according to the 
absolute value of the CI coefficient. 
The converged results of AFQMC/MD and AFQMC/VAFQMC (in their respective limit of large $P$) are shown by a thick line shaded in the corresponding color, with the line width indicating statistical errors
(b) The computational accuracy and efficiency of 
AFQMC/VAFQMC as a function of $P$.
The statistical accuracy is shown as stars (left scale), with the dashed orange curve to guide the eye. The computational cost (normalized to 
that of $P=5$) is shown as squares (right scale).
}
\label{Fig.N2_MD} 
\end{figure}

In Fig.~\ref{Fig.N2_MD}, we take again the worst case scenario in
N$_2$, 
at $R=3.6\,$Bohr, to illustrate 
the computational efficiency and scaling of the method.
In panel (a), we show the energy convergence with respect to $P$.
AFQMC/VAFQMC with $P=5$ yields a slightly worse result than the final converged answer 
for this wave function, but it essentially reaches chemical accuracy already. 
In the inset, we show results from AFQMC calculations with multi-determinants (MD) truncated from a complete active space self-consistent field (CASSCF) calculation,
with $12$ active 
orbitals and $5$ active 
electrons. We then truncate the CASSCF wave function according to the absolute value of the CI configuration, with  
cut-off: $0.1, 0.01, 0.001$, and $0.0005$, respectively.  
We see that the AFQMC/MD calculation requires $\sim 6000$ determinants to reach convergence. It should be noted that there is a great deal of speedup available, because of the structrue of a CI trial wave function~\cite{Hao_Some_recent_developments, Morales_MD_AFQMC, SHCI_AFQMC}, to reduce the computational cost of the AFQMC/MD calculations. Nevertheless, the very large reduction in the number of determinants that the new sampling algorithm allows is significant.

In panel (b), we show the behavior of statistical error bar and related computational cost as a function of $P$ in the AFQMC/VAFQMC algorithm.
All calculations here are performed with the same parameters except for different $P$. As $P$ increases, we observe a decrease of the statistical error, with a linear increase in computational cost. 

\begin{figure}[htbp]
\includegraphics[scale = 0.2]{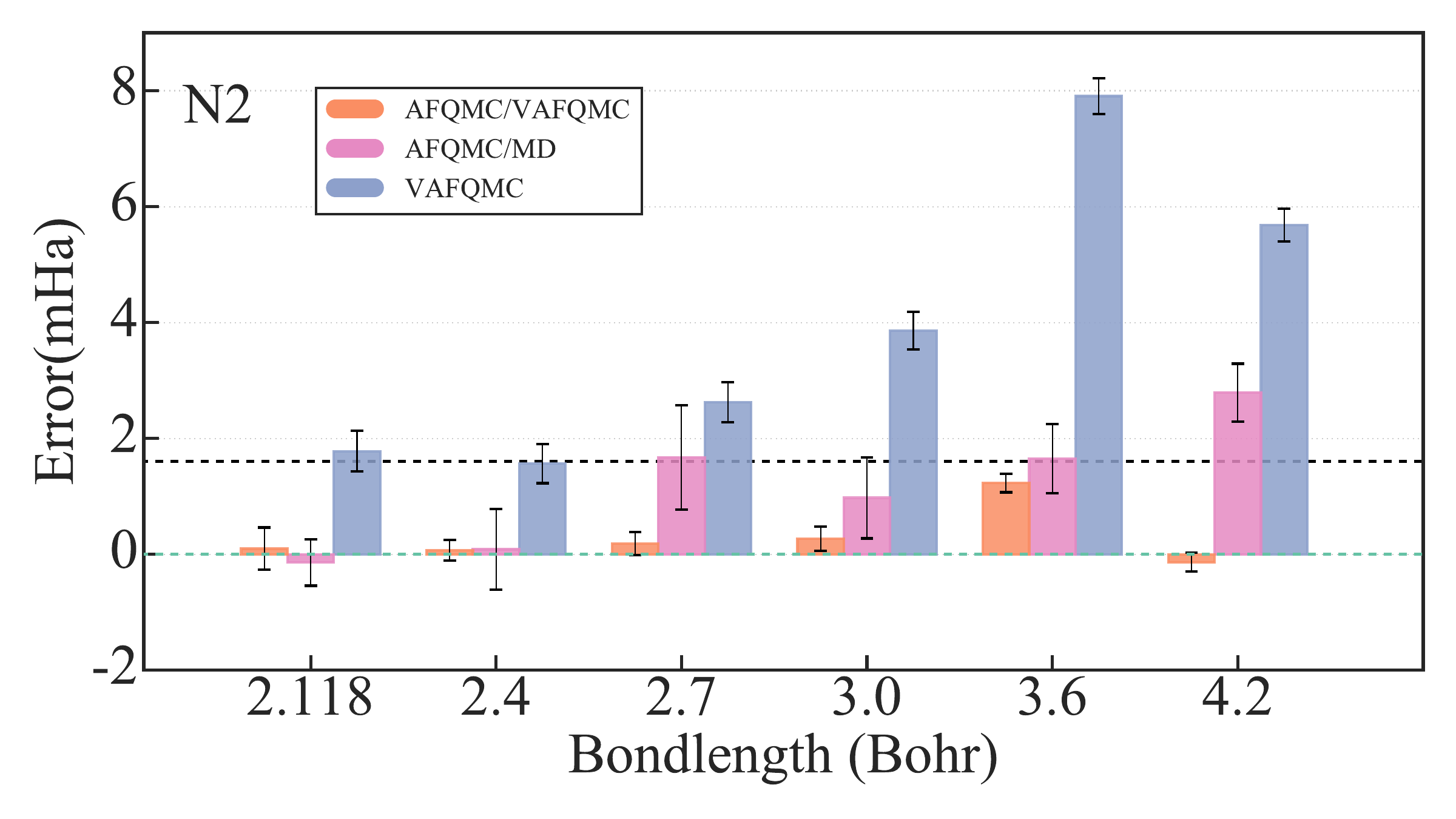} 
\caption{Deviations in the computed ground-state energy from VAFQMC and AFQMC/VAFQMC, for the N$_2$ molecule at different bondlengths, in the cc-pVDZ basis. 
Energies are compared with near-exact DMRG results 
\cite{N2_bondbreaking_DMRG, AFQMC_bondBreaking}.
AFQMC/MD, with $|\Psi_T\rangle$ from truncated MCSCF, was taken from 
Ref.~\cite{AFQMC_bondBreaking} and shown for reference.
The dashed black line represents the chemical accuracy (1\,kcal/mol).} 
\label{Fig.N2} 
\end{figure}

In Fig.~\ref{Fig.N2}, we summarize the ground state energy computed by AFQMC/VAFQMC 
for different $R$ beyond the equilibrium bondlength $R_{\rm eq}=2.118\,$Bohr. 
The variational energy from our trial wave function is shown as VAFQMC. For reference, the AFQMC/MD result is also shown, and MD is selected from multi-configurational self-consistent field (MCSCF) with weight cutoﬀ $0.01$.  
All AFQMC/VAFQMC calculations were 
run for $24$ hours to reach a statistical error of around $0.2$\,mHa. (The run time can be reduced to less than $1$ hour if we target a statistical accuracy of  $1$\,mHa, or roughly chemical accuracy.)
Additional optimization can further reduce the computational cost in AFQMC with Metropolis constraint, which we will leave for 
future work. 

\subsection{\label{sec:results_CuO_FeO}Transition metal oxide molecules}
This section focuses on transition metal oxide molecules, which are prototypical systems exhibiting strong correlations 
that illustrate some of the challenges with standard quantum
chemistry methods. A number of benchmark studies involving AFQMC exist on these molecules~\cite{James-IP,simons_material_2020}. 
Here we directly compare with the near-exact semistochastic heat-bath configuration interaction (SHCI) energies from Ref.~\cite{simons_material_2020}, using identical settings (bondlengths, pseudo-potential, and basis sets). 

\begin{figure}[htbp]
\includegraphics[scale = 0.2]{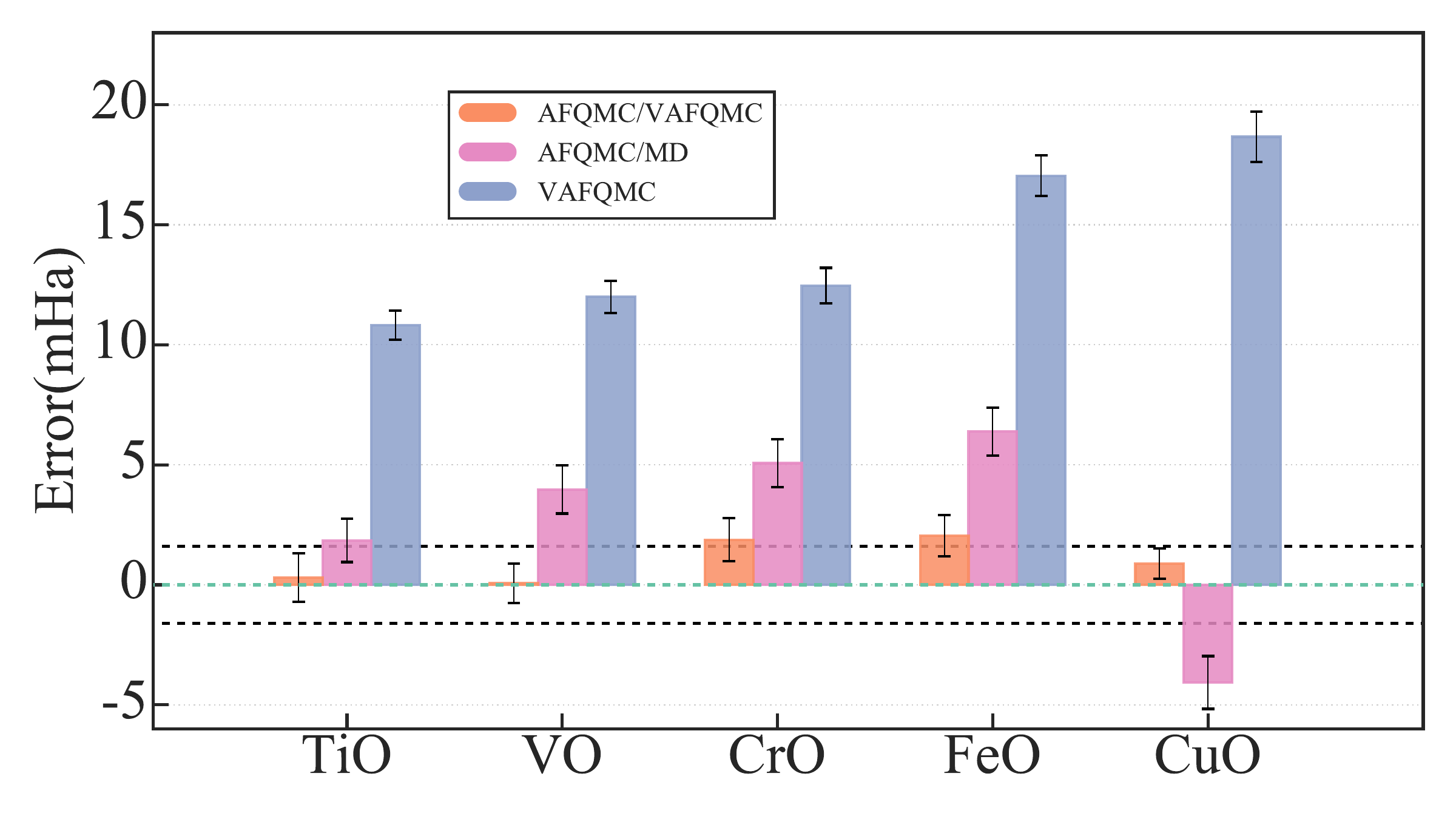} 
\caption{Energy deviations for VAFQMC and AFQMC on five transition metal diatomics.  
Calculations are performed in the vdz basis, with identical setup as in Ref.~\cite{simons_material_2020},
where the reference SHCI energies and AFQMC/MD results are taken 
The dashed black line represents chemical accuracy (1\,kcal/mol).} 
\label{Fig.CuO_FeO} 
\end{figure}

In Fig.~\ref{Fig.CuO_FeO}, we summarize the ground state energy measured by AFQMC with VAFQMC trial wave functions for five diatomics. 
VAFQMC is executed to provide trial wave functions of the form in Eq.~(\ref{eq:VAFQMC}). The variational energies of the VAFQMC 
$|\Psi_T\rangle$ are 
presented in the figure. It is seen that 
these single-time-slice trial wave functions are in themselves limited, giving energy errors as large as $\sim 20$\,mHa. 
AFQMC using these trial wave functions, with 
Metropolis sampling, yields results 
systematically within chemical accuracy, 
indicated by the dashed line). In these calculations, we used 
$P=20$ paths which was verified to be sufficient for sampling
$|\Psi_T\rangle$ (see study in Sec.~\ref{sec:results_N2}). 
Our AFQMC/VAFQMC calculations were executed for 24 hours
(using 240 cores as mentioned earlier), resulting in statistical errors of approximately $0.9$ mHa.

For comparison, we have also included in Fig.~\ref{Fig.CuO_FeO} the AFQMC/MD results from Ref.~\cite{simons_material_2020}.
Similar to what is seen in N$_2$, the new method outperforms AFQMC/MD with multiple-determinant CI-type trial wave functions, in this case obtained from 
truncated CASSCF. 
As mentioned earlier, these CI-type multi-determinant $|\Psi_T\rangle$'s lend themselves to speedup algorithms which 
allow a very large number of determinants, so these AFQMC/MD results can be improved without necessarily incurring high cost. So direct comparisons of the number of determinants is not very meaningful. However, the MD trial wave functions used in these were reasonably advanced, with ${\mathcal O}(1000)$ 
determinants.
That simple one-slice VAFQMC trial wave functions can lead to such a clear improvement over them is thus very encouraging.

\section{\label{sec:summary} Summary and discussion}

This work introduces a way to sample advanced many-body trial wave functions in fermion quantum Monte Carlo when the trial wave function cannot be computed straightforwardly. We have in particular focused on
a scalable algorithm for AFQMC to implement 
a wide class of many-body wave functions as trial wave function. This is accomplished by 
interfacing the branching random walks of AFQMC with  Metropolis sampling of the trial wave function. 
We discussed the mathematical and theoretical basis for the method,
and outlined the algorithm to execute it. Test results are presented in atoms and molecules, including strongly correlated molecules under bond stretching or containing transition metal elements. The method consistently reaches chemical accuracy and demonstrates good  scaling with moderate computational cost. 

Key elements in the new method include the mitigation of systematic errors 
by computing overlap ratios with correlated samples and keeping the trial wave functions tethered to the random walkers to avoid or minimize re-equilibration in the MCMC. 
Without these ingredients it would be impractical to achieve an efficient implementation of any stochastic sampling approach, including proposals on quantum computers. 
Even in an AFQMC calculation of modest system size, 
the overlap itself can vary by 50 orders of magnitude \cite{Hao_infinite_variance}.
Any calculation based on direct evaluations of the overlap itself 
would have fundamental difficulty
to scale up properly with system size. 
Similarly, without the ability to generate samples of $|\Psi_T\rangle$ with importance sampling tethered to the walker, it would be difficult to maintain fidelity of the constraint due to large fluctuations from random sampling.

Our results provide interesting insight to the behavior of noisy constraints in AFQMC.
Even with very modest number of samples, 
with proper importance sampling via tethering 
and the leapfrog approach, a rather modest number of samples ($P\sim 5$) yield good fidelity and statistical accuracy in a realistic molecular calculation, as shown in Fig.~\ref{Fig.N2_MD}. These  observations suggest a promising prospect for the efficient implementation of noisy constraints in AFQMC, provided efficient sampling schemes. 

Though our algorithm is presented for trial wave functions in the form of Eq.~\ref{eq:trial}, any wave function ansatz can be implemented to AFQMC as constraint if its overlap with given determinants can be evaluated effectively through Monte Carlo sampling. 
While we have formulated the trial wave function as integrals over 
auxiliary-fields, this is certainly not a requirement. 
For example, to date applications of CI-like trial wave functions, such as the
truncated MCSCF wave function illustrated in Sec.~\ref{sec:Results},
take a large number of determinants explicitly. 
Instead of using a deterministic, preset sequence, 
one could imagine approximating these wave functions with a small number of samples and incorporating them into each propagation/measurement procedure.
It could also be useful to apply a $|\Psi_T\rangle$ derived from a density matrix renormalization group (DMRG) or tensor network calculations, by sampling it 
in CI space \cite{MPS_AFQMC}.
Additionally, other forms of variational Monte Carlo (VMC) or neural network 
wave functions could also be considered. Indeed the single-time-slice trial wave function 
we have considered in this work for N$_2$ and transition metal diatomics are very similar to a Slater-Jastrow wave function which is traditional in VMC. (Our VAFQMC contains beyond density-density correlations and are thus more general.) 

Our algorithm as currently implemented scales proportionally to the number of Metropolis samples (roughly $P\times M$ ), essentially giving a prefactor to the cost of a standard AFQMC with single determinant trial wave function. As mentioned, the number of paths tethered to 
each walker is of ${\mathcal O}(10)$ and the number of update sweeps is of ${\mathcal O}(1)$. Since the quality of the trial wave function actually leads to a reduction in the variance (as seen in Fig.~\ref{Fig.N2_MD}), the additional cost has not been too onerous. 
As a first demonstration of the approach for AFQMC with stochastically sampled constraint, our implementation here has not really focused on speedups, acceleration, or 
computational efficiency. Opportunities clearly exist to optimize and improve. It will be valuable to incorporate many of the ideas which have lead to great efficiency gain and optimization in deterministic trial wave functions~\cite{Morales_MD_AFQMC,
SHCI_AFQMC, Ankit_CISD,Hao_Some_recent_developments}. Additionally, we can optimize with respect to the interplay between time step size $\tau$ and how to optimally mix and match step 1 and step 2 in the leapfrog update.

Interfacing BRW with Metropolis is not limited to ground-state AFQMC. Extension to finite temperature AFQMC \cite{zhang1999finite, Yuan-Yao_FT_AFQMC} is straightforward. This method is also not limited to auxiliary-field space. It can similarly be applied in coordinate-space QMC, for example, interfacing diffusion Monte Carlo (DMC) with path-integral methods or some other form to sample a VMC/neural network wave function. 

\section{\label{sec:level7} Acknowledgments }
Authors thank Ryan Levy and Kang Wang for providing useful data. 
Z.X. is grateful for the support and hospitality of the Center for Computational Quantum Physics (CCQ) during early part of this work. 
The Flatiron Institute is a division of the Simons Foundation.
Computing was carried out through the support from Institute of Physics, Chinese Academy of Sciences.

\bibliography{cite} 

\section{\label{sec:APPENDIX}APPENDIX}

\subsection{\label{sec:APPENDIX_VAFQMC}Summary of VAFQMC}
We briefly discuss the VAFQMC ansatz used in this work. For more details please refer to Ref.~\cite{HAFQMC}.
VAFQMC method variationally optimizes a finite-time propagator in AFQMC to approximate the ground state. It allows the propagator to vary at different imaginary times and optimizes the initial Slater determinant, leading to the variational ansatz:
\begin{equation}
    \ket{\psi_\theta} = \exp\bqty{ \int_0^\tau -\tilde{H}(s) \dd s } \ket{\tilde{\psi}_I},
\end{equation}
where $\theta$ denotes all parameters. After time discretization and the Hubbard-Stratonovich transformation, the ansatz becomes:
\begin{equation}
\label{eq:ansatz}
\begin{split}
    \ket{\psi_\theta}
    =&  \int { \prod_{\gamma,l} \frac{\dd{x_{\gamma l}}}{\sqrt{2\pi}}  e^{-\frac{1}{2}x_{\gamma l}^2} }  \\
    &   { \prod_l^{N_l} \exp\bqty{-t_l\tilde{T}_l} \exp\bqty{\sqrt{-s_l} \sum_\gamma^{N_\gamma}  x_{\gamma l} {\tilde{L}_\gamma}}} \\
    &   \exp\bqty{-t_0\tilde{T}_0} \ket{\tilde{\psi}_I} \\
    \equiv& \int \dd{\vx} \tilde{U}(\vx) \ket{\tilde{\psi}_I},
\end{split}
\end{equation}
where $t_l$ and $s_l$ are optimizable step sizes, and $\vx = \Bqty{x_{\gamma l}}$. The sign problem is addressed practically by using fewer projection steps and a soft restraint. Since the ansatz is optimized variationally, we no longer suffer from Trotter error, which allows us to use larger time steps and lower order Taylor expansion of the matrix exponential. In this work, we limit the ansatz to a single step ($l=1$) to further speedup the optimization procedure. 

The optimization of the ansatz is done by minimizing the energy expectation value:
\begin{equation}
    \label{eq:eraw}
    E_\theta
    = \frac{\ev{\hH}{\psi_\theta}}{\braket{\psi_\theta}}
    = \frac{ \int \dd{\vx}\dd{\vx'} \ev{\tilde{U}^\dag(\vx') \hH \tilde{U}(\vx)}{\tilde{\psi}_I} }{ \int \dd{\vx}\dd{\vx'} \ev{\tilde{U}^\dag(\vx') \tilde{U}(\vx)}{\tilde{\psi}_I} },
\end{equation}
which is computed via Monte Carlo by separating the overlap amplitude into magnitude $\cD$ and phase $\cS$, and defining the local energy $\cE$:
\begin{align}
    \cD(\vx, \vx') & = \abs{\ev{\tilde{U}^\dag(\vx') \tilde{U}(\vx)}{\tilde{\psi}_I}} \, ,\\
    \cS(\vx, \vx') & = \frac{\ev{\tilde{U}^\dag(\vx') \tilde{U}(\vx)}{\tilde{\psi}_I}}{\abs{\ev{\tilde{U}^\dag(\vx') \tilde{U}(\vx)}{\tilde{\psi}_I}}} \, , \\
    \cE(\vx, \vx') & = \frac{ \ev{\tilde{U}^\dag(\vx') \hH \tilde{U}(\vx)}{\tilde{\psi}_I} }{ \ev{\tilde{U}^\dag(\vx') \tilde{U}(\vx)}{\tilde{\psi}_I} } \, .
\end{align}
The energy is then given by:
\begin{equation}
    \label{eq:esym}
    E_\theta
    = \frac{\int \dd{\vx}\dd{\vx'} \cD(\vx, \vx') \cE(\vx, \vx') \cS(\vx, \vx') }{ \int \dd{\vx}\dd{\vx'} \cD(\vx, \vx') \cS(\vx, \vx') }
    = \frac{ \ev{\cE \cS}_\cD }{ \ev{\cS}_\cD }.
\end{equation}
A one-sided restraint on the average sign $S_\theta = \Re \bqty{\ev{\cS}_\cD}$ is used to mitigate the sign problem:
\begin{equation}
    \label{eq:min}
    \min_\theta \bqty{ E_\theta + \lambda\, \pqty{\mathrm{max}\Bqty{B - S_\theta, 0}}^2 },
\end{equation}
where we take hyperparameters $\lambda=1$ and $B=0.7$ . This restraint avoids systematic bias and maintains a sufficient average sign.

\subsection{\label{sec:APPENDIX_summary}Summary for AFQMC with Metropolis constraint}
\begin{figure*}[htbp]
\includegraphics[scale = 0.27]{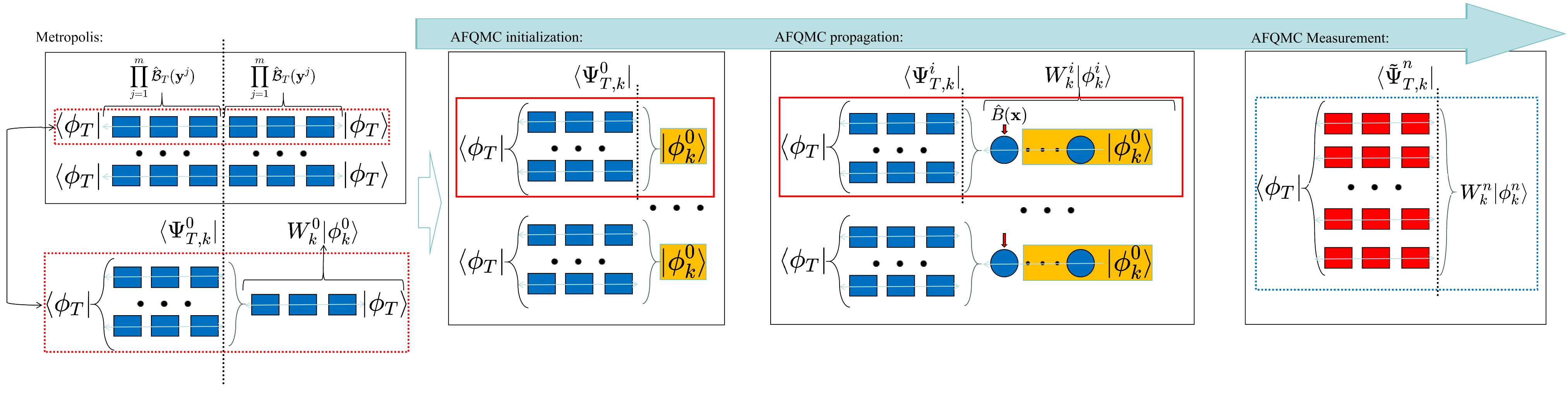} 
\caption{
AFQMC with a stochastically sampled trial wave function. 
Filled 
rectangles/circles denote Metropolis/BRW sampled operators. 
Monte Carlo sampling is 
performed on objects highlighted 
with dashed (Metropolis) or solid (BRW) rectangular boxes.
The panel ``Metropolis" on the left 
shows the estimation of trial energy $\frac{\langle \Psi_T|\hat{H}|\Psi_T\rangle}{\langle \Psi_T|\Psi_T\rangle}$ with sampled Metropolis paths. 
The Metropolis paths can be reformulated to form the composite walker (i.e., in the bottom dashed red rectangle) to 
initialize AFQMC, as the panel ``AFQMC initialization" shows.
In ``AFQMC propagation", 
the composite walker 
is propagated by inserting operators $\hat{B}(\textbf{x})$ 
from right to left as the red arrow demonstrates, through the ``leap-frog" way demonstrated in Fig.~\ref{Fig.AFQMC_Metro}.
The auxiliary fields $\textbf{x}$ are fixed once sampled by BRW (i.e., absorbed into determinants implied by ``orange background") while auxiliary fields $\textbf{y}$ are updated by Metropolis continuously. 
During ``AFQMC measurement", 
more Metropolis samples can be sampled 
to provide a better estimation of $\langle \Psi_T|$ (i.e., $\langle \tilde{\Psi}^n_{T,k}|$ with filled red blocks) than those in the propagation to facilitate the numerical efficiency.
} 
\label{Fig.summary} 
\end{figure*}

In this section, we elaborate on the details of implementing AFQMC with Metropolis constraint. Main procedures are summarized in  Fig.~\ref{Fig.summary}, which can be divided into three main steps: initialization, propagation, and measurement.

\subsubsection{Initialization}
The implementation of AFQMC with Metropolis constraint can start from a standard Metropolis algorithm that estimates $\langle\hat{O} \rangle$ with trial wave function Eq.~\ref{eq:trial}  through
\begin{equation}
\begin{aligned}
\langle\hat{O} \rangle_T=\frac{\langle \Psi_T|\hat{O}|\Psi_T\rangle}{\langle \Psi_T|\Psi_T\rangle} \overset{\mathrm{Metro}}{\longrightarrow}\frac{\sum_{k}\sum_{k'}e^{i\theta_{k',k}}\langle O \rangle_{k', k}}{\sum_{k}\sum_{k'}e^{i\theta_{k',k}}}
\label{eq:Metropolis}
\end{aligned}
\end{equation}
where we define a local estimation of $\hat{O}$ according to a given sample of auxiliary fields (i.e., $Y_k$, $Y_{k'}$) that are labeled by $k$ and $k'$ respectively:
\begin{equation}
\begin{aligned}
\langle O \rangle_{k', k}=\frac{\langle\phi_T|p_T(Y_{k'})  \hat{\mathcal B}_T(Y_{k'})\,\hat{O}\,p_T(Y_{k})\hat{\mathcal B}_T(Y_{k})|\phi_T\rangle}{\langle\phi_T|p_T(Y_{k'})\hat{\mathcal B}_T(Y_{k'})\,p_T(Y_{k})\hat{\mathcal B}_T(Y_{k})|\phi_T\rangle}
\label{eq:Metropolis_chain}
\end{aligned}
\end{equation}
and complex phase
$$
e^{i\theta_{k',k}} = \frac{\langle\phi_T|\hat{\mathcal B}_T(Y_{k'})\hat{\mathcal B}_T(Y_{k})|\phi_T\rangle}{|\langle\phi_T|\hat{\mathcal B}_T(Y_{k'})\hat{\mathcal B}_T(Y_{k})|\phi_T\rangle|}.
$$
As presented in Eq.~\ref{eq:Metropolis}, the Metropolis estimation of $\langle\hat{O} \rangle$ can be understood as a weighted average of $\langle O \rangle_{k', k}$ with weight $e^{i\theta_{k',k}}$.

To interface BRW with Metropolis, we initialize
\begin{equation}
\begin{aligned}
\langle \Psi^{0}_{T,k}| = \sum^P_{p=1} \frac{\langle \phi_T|\hat{\mathcal B}_T(Y[p]) }{|\langle \phi_T|\hat{\mathcal B}_T(Y[p]) |\phi^0_{k} \rangle|},
\end{aligned}
\end{equation}
with 
\begin{equation}
\begin{aligned}
|\phi^0_k \rangle = \hat{\mathcal B}_T(Y_k) |\phi_T\rangle,
\end{aligned}
\end{equation}
and change notations $k' \rightarrow p$ and $\hat{\mathcal B}_T(Y_{k'}) \rightarrow \hat{\mathcal B}_T(Y[p])$ to consist with Eq.~\ref{eq:Metropolis_sampled_trial} where $p$ labels the path tethered to walker $k$. It is straightforward to see that the Metropolis sampling Eq.~\ref{eq:Metropolis} can be smoothly adopted to the initialization of AFQMC:
\begin{equation}
\begin{aligned}
\langle\hat{O} \rangle_T \overset{\mathrm{AFQMC}}{\longrightarrow}\frac{\sum_{k} W^0_k\frac{\langle \Psi^0_{T,k} |\hat{H} |\phi^{0}_k\rangle}{\langle \Psi^0_{T,k}|\phi^{n}_k\rangle} }{\sum_{k}W^0_k}
\label{eq:AFQMC_Metropolis_initial}
\end{aligned}
\end{equation}
with 
$$
W^0_k = \sum_{k'}e^{i\theta_{k',k}}.
$$

\subsubsection{Propagation}
Basic theories for the propagation are discussed in Sec.~\ref{sec:results_method}. Here, we provide 
the 
implementation details in the form of 
a pseudo-code: 
\begin{enumerate}
    \item BRW advance walkers to $| \phi^{i+1}_{k}\rangle$ from $| \phi^{i}_{k}\rangle$ according to $\langle \Psi^{i}_{T,k}|$: 
    \begin{itemize}
        \item sample a field $\textbf{x}$ with probability $p(\textbf{x})$\,,
        \item evaluate dynamic shift
        $$
        \overline{\textbf{x}}^i_{k,\gamma} =-\sqrt{\tau}\frac{\langle  \Psi^i_{T,k}|\hat{L}_\gamma |\phi^i_k \rangle}{\langle \Psi^i_{T,k}| \phi^i_k\rangle}
        $$
        and
        $$
        I(\textbf{x}, \overline{\textbf{x}}^i_k, \phi^i_k) = \frac{p(\textbf{x}-\overline{\textbf{x}}^i_k)}{p(\textbf{x})}\frac{\langle \Psi^i_{T,k}| \hat{B}(\textbf{x} - \overline{\textbf{x}}^i_k)| \phi^i_k \rangle }{\langle \Psi^i_{T,k}|\phi^i_k \rangle }\,.
        $$
        \item advance walker
        $$
        | \phi_k^{i+1} \rangle = \hat{B}(\textbf{x} - \overline{\textbf{x}}_k^i)| \phi_k^i \rangle
        $$
        \item assign weight $W^{i+1}_k$ under the phaseless approximation:
        $$
        W_k^{i+1} = W^i_k * \bigr|I(\textbf{x}, \overline{\textbf{x}}_k^i, \phi_k^i)\bigr| * \mathrm{max}[0,\cos(\mathrm{Arg}\frac{\langle \Psi^i_{T,k}| \phi^{i+1}_k \rangle }{\langle \Psi^i_{T,k}|\phi^{i}_k \rangle})].
        $$
    \end{itemize} 
    \item Metropolis update $Y[p]$ according to $| \phi^{i+1}_{k}\rangle$: 
    \begin{itemize}
        \item sample $Y'[p]$ with probability $p(Y'[p])$. 
        \item accept sampled $Y'[p]$ according to ratio 
            $$
            \alpha = \frac{ |\langle \phi_T|\hat{\mathcal B}_T(Y'[p]) |\phi^{i+1}_k\rangle|}{ |\langle \phi_T|\hat{\mathcal B}_T(Y[p]) |\phi^{i+1}_k\rangle|}.
            $$
            If $Y'[p]$ is accepted, set $Y[p]=Y'[p]$ or set $Y'[p]=Y[p]$. 
        \item iterate above to reach equilibrium.
    \end{itemize}   
    \item Evaluate $\Delta\theta_T$ and update weight $W_k^{i+1} = W_k^{i+1}*\mathrm{max}[0,\cos(\Delta\theta_T)]$ according to Eq.~\ref{eq:AFQMC_Metro_constraint_cos}.
\end{enumerate}

\subsubsection{Measurement}
Suppose we restore the last $m$ projectors $\hat{B}(\textbf{x})$ in our propagation, the measurement of observables $\hat{O}$, including dynamic correlations \cite{Ettore_dynamic_correlation} that do not commute with Hamiltonian, can be achieved by inserting related operators directly into the middle of sampled Metropolis paths that determined by $(Y_k,X_k)$:
\begin{equation}
\begin{aligned}
\frac{\langle \Psi_T|[e^{-\hat{H}\tau}]^{\frac{m}{2}}\hat{O}[e^{-\hat{H}\tau}]^{n+\frac{m}{2}}|\Psi_T\rangle}{\langle \Psi_T|[e^{-\hat{H}\tau}]^{\frac{m}{2}}[e^{-\hat{H}\tau}]^{n+\frac{m}{2}}|\Psi_T\rangle} \!\!\overset{\mathrm{AFQMC}}{\longrightarrow}\!\! \frac{\sum_k W^{n+m}_k \langle O \rangle^{\frac{m}{2},n+\frac{m}{2}}_{k}}{\sum_{k} W^{n+m}_{k}}
\label{eq:AFQMC_Metropolis_measure}
\end{aligned}
\end{equation}
with 
\begin{equation}
\begin{aligned}
\langle O \rangle^{\frac{m}{2},n+\frac{m}{2}}_{k} \!\!=\!\! \frac{\langle \Psi^{n+m}_{T,k}|[\prod^{n+m}_{i=n+\frac{m}{2}}\hat{B}(\textbf{x}^{i}_k)]\hat{O}[\prod^{n+\frac{m}{2}}_{i=n+1}\hat{B}(\textbf{x}^{i}_k)]|\phi^n_k\rangle}{\langle \Psi^{n+m}_{T,k}|\prod^{n+m}_{i=n+1}\hat{B}(\textbf{x}^{i}_k)|\phi^{n}_k\rangle}.
\label{eq:AFQMC_Metropolis_measure2}
\end{aligned}
\end{equation}

When doing the measurement described in Eq.~\ref{eq:AFQMC_Metropolis_measure}, we can again apply Metropolis update on auxiliary fields $Y[p]$ in $\langle \Psi^{n+m}_{T,k}|$ to sample a set of new auxiliary fields $Y'[p]$ and related $\langle \tilde{\Psi}^{n+m}_{T,k}|$. The measurements in Eq.~\ref{eq:AFQMC_Metropolis_measure} then can be updated through
\begin{equation}
\begin{aligned}
\langle \Psi^{n+m}_{T,k}| \rightarrow \langle \tilde{\Psi}^{n+m}_{T,k}|
\end{aligned}
\end{equation}
and
\begin{equation}
\begin{aligned}
W^{n+m}_k \rightarrow W^{n+m}_k*\frac{\sum_{p'=1}^{P'} S(Y'[p'])}{\sum_{p=1}^P S(Y[p])}
\end{aligned}
\end{equation}
with $ S(Y'[p'])$ and $S(Y[p])$ to specify related sign ratio. 

\end{document}